\documentstyle{aipproc2}
\begin{document}

\catcode`!=11 

  

\def\PiC{P\kern-.12em\lower.5ex\hbox{I}\kern-.075emC}
\def\PiCTeX{\PiC\kern-.11em\TeX}

\def\!ifnextchar#1#2#3{%
  \let\!testchar=#1%
  \def\!first{#2}%
  \def\!second{#3}%
  \futurelet\!nextchar\!testnext}
\def\!testnext{%
  \ifx \!nextchar \!spacetoken 
    \let\!next=\!skipspacetestagain
  \else
    \ifx \!nextchar \!testchar
      \let\!next=\!first
    \else 
      \let\!next=\!second 
    \fi 
  \fi
  \!next}
\def\\{\!skipspacetestagain} 
  \expandafter\def\\ {\futurelet\!nextchar\!testnext} 
\def\\{\let\!spacetoken= } \\  

\def\!tfor#1:=#2\do#3{%
  \edef\!fortemp{#2}%
  \ifx\!fortemp\!empty 
    \else
    \!tforloop#2\!nil\!nil\!!#1{#3}%
  \fi}
\def\!tforloop#1#2\!!#3#4{%
  \def#3{#1}%
  \ifx #3\!nnil
    \let\!nextwhile=\!fornoop
  \else
    #4\relax
    \let\!nextwhile=\!tforloop
  \fi 
  \!nextwhile#2\!!#3{#4}}

\def\!etfor#1:=#2\do#3{%
  \def\!!tfor{\!tfor#1:=}%
  \edef\!!!tfor{#2}%
  \expandafter\!!tfor\!!!tfor\do{#3}}

\def\!cfor#1:=#2\do#3{%
  \edef\!fortemp{#2}%
  \ifx\!fortemp\!empty 
  \else
    \!cforloop#2,\!nil,\!nil\!!#1{#3}%
  \fi}
\def\!cforloop#1,#2\!!#3#4{%
  \def#3{#1}%
  \ifx #3\!nnil
    \let\!nextwhile=\!fornoop 
  \else
    #4\relax
    \let\!nextwhile=\!cforloop
  \fi
  \!nextwhile#2\!!#3{#4}}

\def\!ecfor#1:=#2\do#3{%
  \def\!!cfor{\!cfor#1:=}%
  \edef\!!!cfor{#2}%
  \expandafter\!!cfor\!!!cfor\do{#3}}

\def\!empty{}
\def\!nnil{\!nil}
\def\!fornoop#1\!!#2#3{}

\def\!ifempty#1#2#3{%
  \edef\!emptyarg{#1}%
  \ifx\!emptyarg\!empty
    #2%
  \else
    #3%
  \fi}

\def\!getnext#1\from#2{%
  \expandafter\!gnext#2\!#1#2}%
\def\!gnext\\#1#2\!#3#4{%
  \def#3{#1}%
  \def#4{#2\\{#1}}%
  \ignorespaces}

%
\def\!getnextvalueof#1\from#2{%
  \expandafter\!gnextv#2\!#1#2}%
\def\!gnextv\\#1#2\!#3#4{%
  #3=#1%
  \def#4{#2\\{#1}}%
  \ignorespaces}

\def\!copylist#1\to#2{%
  \expandafter\!!copylist#1\!#2}
\def\!!copylist#1\!#2{%
  \def#2{#1}\ignorespaces}

\def\!wlet#1=#2{%
  \let#1=#2 
  \wlog{\string#1=\string#2}}

\def\!listaddon#1#2{%
  \expandafter\!!listaddon#2\!{#1}#2}
\def\!!listaddon#1\!#2#3{%
  \def#3{#1\\#2}}


\def\!rightappend#1\withCS#2\to#3{\expandafter\!!rightappend#3\!#2{#1}#3}
\def\!!rightappend#1\!#2#3#4{\def#4{#1#2{#3}}}

\def\!leftappend#1\withCS#2\to#3{\expandafter\!!leftappend#3\!#2{#1}#3}
\def\!!leftappend#1\!#2#3#4{\def#4{#2{#3}#1}}

\def\!lop#1\to#2{\expandafter\!!lop#1\!#1#2}
\def\!!lop\\#1#2\!#3#4{\def#4{#1}\def#3{#2}}



\def\!loop#1\repeat{\def\!body{#1}\!iterate}
\def\!iterate{\!body\let\!next=\!iterate\else\let\!next=\relax\fi\!next}

\def\!!loop#1\repeat{\def\!!body{#1}\!!iterate}
\def\!!iterate{\!!body\let\!!next=\!!iterate\else\let\!!next=\relax\fi\!!next}

\def\!removept#1#2{\edef#2{\expandafter\!!removePT\the#1}}
{\catcode`p=12 \catcode`t=12 \gdef\!!removePT#1pt{#1}}

\def\placevalueinpts of <#1> in #2 {%
  \!removept{#1}{#2}}

\def\!mlap#1{\hbox to 0pt{\hss#1\hss}}
\def\!vmlap#1{\vbox to 0pt{\vss#1\vss}}

\def\!not#1{%
  #1\relax
    \!switchfalse
  \else
    \!switchtrue
  \fi
  \if!switch
  \ignorespaces}




\let\!!!wlog=\wlog              
\def\wlog#1{}    

\newdimen\headingtoplotskip     
\newdimen\linethickness         
\newdimen\longticklength        
\newdimen\plotsymbolspacing     
\newdimen\shortticklength       
\newdimen\stackleading          
\newdimen\tickstovaluesleading  
\newdimen\totalarclength        
\newdimen\valuestolabelleading  

\newbox\!boxA                   
\newbox\!boxB                   
\newbox\!picbox                 
\newbox\!plotsymbol             
\newbox\!putobject              
\newbox\!shadesymbol            

\newcount\!countA               
\newcount\!countB               
\newcount\!countC               
\newcount\!countD               
\newcount\!countE               
\newcount\!countF               
\newcount\!countG               
\newcount\!fiftypt              
\newcount\!intervalno           
\newcount\!npoints              
\newcount\!nsegments            
\newcount\!ntemp                
\newcount\!parity               
\newcount\!scalefactor          
\newcount\!tfs                  
\newcount\!tickcase             

\newdimen\!Xleft                
\newdimen\!Xright               
\newdimen\!Xsave                
\newdimen\!Ybot                 
\newdimen\!Ysave                
\newdimen\!Ytop                 
\newdimen\!angle                
\newdimen\!arclength            
\newdimen\!areabloc             
\newdimen\!arealloc             
\newdimen\!arearloc             
\newdimen\!areatloc             
\newdimen\!bshrinkage           
\newdimen\!checkbot             
\newdimen\!checkleft            
\newdimen\!checkright           
\newdimen\!checktop             
\newdimen\!dimenA               
\newdimen\!dimenB               
\newdimen\!dimenC               
\newdimen\!dimenD               
\newdimen\!dimenE               
\newdimen\!dimenF               
\newdimen\!dimenG               
\newdimen\!dimenH               
\newdimen\!dimenI               
\newdimen\!distacross           
\newdimen\!downlength           
\newdimen\!dp                   
\newdimen\!dshade               
\newdimen\!dxpos                
\newdimen\!dxprime              
\newdimen\!dypos                
\newdimen\!dyprime              
\newdimen\!ht                   
\newdimen\!leaderlength         
\newdimen\!lshrinkage           
\newdimen\!midarclength         
\newdimen\!offset               
\newdimen\!plotheadingoffset    
\newdimen\!plotsymbolxshift     
\newdimen\!plotsymbolyshift     
\newdimen\!plotxorigin          
\newdimen\!plotyorigin          
\newdimen\!rootten              
\newdimen\!rshrinkage           
\newdimen\!shadesymbolxshift    
\newdimen\!shadesymbolyshift    
\newdimen\!tenAa                
\newdimen\!tenAc                
\newdimen\!tenAe                
\newdimen\!tshrinkage           
\newdimen\!uplength             
\newdimen\!wd                   
\newdimen\!wmax                 
\newdimen\!wmin                 
\newdimen\!xB                   
\newdimen\!xC                   
\newdimen\!xE                   
\newdimen\!xM                   
\newdimen\!xS                   
\newdimen\!xaxislength          
\newdimen\!xdiff                
\newdimen\!xleft                
\newdimen\!xloc                 
\newdimen\!xorigin              
\newdimen\!xpivot               
\newdimen\!xpos                 
\newdimen\!xprime               
\newdimen\!xright               
\newdimen\!xshade               
\newdimen\!xshift               
\newdimen\!xtemp                
\newdimen\!xunit                
\newdimen\!xxE                  
\newdimen\!xxM                  
\newdimen\!xxS                  
\newdimen\!xxloc                
\newdimen\!yB                   
\newdimen\!yC                   
\newdimen\!yE                   
\newdimen\!yM                   
\newdimen\!yS                   
\newdimen\!yaxislength          
\newdimen\!ybot                 
\newdimen\!ydiff                
\newdimen\!yloc                 
\newdimen\!yorigin              
\newdimen\!ypivot               
\newdimen\!ypos                 
\newdimen\!yprime               
\newdimen\!yshade               
\newdimen\!yshift               
\newdimen\!ytemp                
\newdimen\!ytop                 
\newdimen\!yunit                
\newdimen\!yyE                  
\newdimen\!yyM                  
\newdimen\!yyS                  
\newdimen\!yyloc                
\newdimen\!zpt                  

\newif\if!axisvisible           
\newif\if!gridlinestoo          
\newif\if!keepPO                
\newif\if!placeaxislabel        
\newif\if!switch                
\newif\if!xswitch               

\newtoks\!axisLaBeL             
\newtoks\!keywordtoks           

\newwrite\!replotfile           

\newhelp\!keywordhelp{The keyword mentioned in the error message in unknown. 
Replace NEW KEYWORD in the indicated response by the keyword that 
should have been specified.}    

\!wlet\!!origin=\!xM                   
\!wlet\!!unit=\!uplength               
\!wlet\!Lresiduallength=\!dimenG       
\!wlet\!Rresiduallength=\!dimenF       
\!wlet\!axisLength=\!distacross        
\!wlet\!axisend=\!ydiff                
\!wlet\!axisstart=\!xdiff              
\!wlet\!axisxlevel=\!arclength         
\!wlet\!axisylevel=\!downlength        
\!wlet\!beta=\!dimenE                  
\!wlet\!gamma=\!dimenF                 
\!wlet\!shadexorigin=\!plotxorigin     
\!wlet\!shadeyorigin=\!plotyorigin     
\!wlet\!ticklength=\!xS                
\!wlet\!ticklocation=\!xE              
\!wlet\!ticklocationincr=\!yE          
\!wlet\!tickwidth=\!yS                 
\!wlet\!totalleaderlength=\!dimenE     
\!wlet\!xone=\!xprime                  
\!wlet\!xtwo=\!dxprime                 
\!wlet\!ySsave=\!yM                    
\!wlet\!ybB=\!yB                       
\!wlet\!ybC=\!yC                       
\!wlet\!ybE=\!yE                       
\!wlet\!ybM=\!yM                       
\!wlet\!ybS=\!yS                       
\!wlet\!ybpos=\!yyloc                  
\!wlet\!yone=\!yprime                  
\!wlet\!ytB=\!xB                       
\!wlet\!ytC=\!xC                       
\!wlet\!ytE=\!downlength               
\!wlet\!ytM=\!arclength                
\!wlet\!ytS=\!distacross               
\!wlet\!ytpos=\!xxloc                  
\!wlet\!ytwo=\!dyprime                 

\!zpt=0pt                              
\!xunit=1pt
\!yunit=1pt
\!arearloc=\!xunit
\!areatloc=\!yunit
\!dshade=5pt
\!leaderlength=24in
\!tfs=256                              
\!wmax=5.3pt                           
\!wmin=2.7pt                           
\!xaxislength=\!xunit
\!xpivot=\!zpt
\!yaxislength=\!yunit 
\!ypivot=\!zpt
\plotsymbolspacing=.4pt
  \!dimenA=50pt \!fiftypt=\!dimenA     

\!rootten=3.162278pt                   
\!tenAa=8.690286pt                     
\!tenAc=2.773839pt                     
\!tenAe=2.543275pt                     

\def\!cosrotationangle{1}      
\def\!sinrotationangle{0}      
\def\!xpivotcoord{0}           
\def\!xref{0}                  
\def\!xshadesave{0}            
\def\!ypivotcoord{0}           
\def\!yref{0}                  
\def\!yshadesave{0}            
\def\!zero{0}                  

\let\wlog=\!!!wlog
%
  
\def\normalgraphs{%
  \longticklength=.4\baselineskip
  \shortticklength=.25\baselineskip
  \tickstovaluesleading=.25\baselineskip
  \valuestolabelleading=.8\baselineskip
  \linethickness=.4pt
  \stackleading=.17\baselineskip
  \headingtoplotskip=1.5\baselineskip
  \visibleaxes
  \ticksout
  \nogridlines
  \unloggedticks}
%
\def\setplotarea x from #1 to #2, y from #3 to #4 {%
  \!arealloc=\!M{#1}\!xunit \advance \!arealloc -\!xorigin
  \!areabloc=\!M{#3}\!yunit \advance \!areabloc -\!yorigin
  \!arearloc=\!M{#2}\!xunit \advance \!arearloc -\!xorigin
  \!areatloc=\!M{#4}\!yunit \advance \!areatloc -\!yorigin
  \!initinboundscheck
  \!xaxislength=\!arearloc  \advance\!xaxislength -\!arealloc
  \!yaxislength=\!areatloc  \advance\!yaxislength -\!areabloc
  \!plotheadingoffset=\!zpt
  \!dimenput {{\setbox0=\hbox{}\wd0=\!xaxislength\ht0=\!yaxislength\box0}}
     [bl] (\!arealloc,\!areabloc)}
%
\def\visibleaxes{%
  \def\!axisvisibility{\!axisvisibletrue}}
\def\invisibleaxes{%
  \def\!axisvisibility{\!axisvisiblefalse}}
%

\def\!fixkeyword#1{%
  \errhelp=\!keywordhelp
  \errmessage{Unrecognized keyword `#1': \the\!keywordtoks{NEW KEYWORD}'}}

\!keywordtoks={enter `i\fixkeyword}

\def\fixkeyword#1{%
  \!nextkeyword#1 }


\def\axis {%
  \def\!nextkeyword##1 {%
    \expandafter\ifx\csname !axis##1\endcsname \relax
      \def\!next{\!fixkeyword{##1}}%
    \else
      \def\!next{\csname !axis##1\endcsname}%
    \fi
    \!next}%
  \!offset=\!zpt
  \!axisvisibility
  \!placeaxislabelfalse
  \!nextkeyword}

\def\!axisbottom{%
  \!axisylevel=\!areabloc
  \def\!tickxsign{0}%
  \def\!tickysign{-}%
  \def\!axissetup{\!axisxsetup}%
  \def\!axislabeltbrl{t}%
  \!nextkeyword}

\def\!axistop{%
  \!axisylevel=\!areatloc
  \def\!tickxsign{0}%
  \def\!tickysign{+}%
  \def\!axissetup{\!axisxsetup}%
  \def\!axislabeltbrl{b}%
  \!nextkeyword}

\def\!axisleft{%
  \!axisxlevel=\!arealloc
  \def\!tickxsign{-}%
  \def\!tickysign{0}%
  \def\!axissetup{\!axisysetup}%
  \def\!axislabeltbrl{r}%
  \!nextkeyword}

\def\!axisright{%
  \!axisxlevel=\!arearloc
  \def\!tickxsign{+}%
  \def\!tickysign{0}%
  \def\!axissetup{\!axisysetup}%
  \def\!axislabeltbrl{l}%
  \!nextkeyword}

\def\!axisshiftedto#1=#2 {%
  \if 0\!tickxsign
    \!axisylevel=\!M{#2}\!yunit
    \advance\!axisylevel -\!yorigin
  \else
    \!axisxlevel=\!M{#2}\!xunit
    \advance\!axisxlevel -\!xorigin
  \fi
  \!nextkeyword}

\def\!axisvisible{%
  \!axisvisibletrue  
  \!nextkeyword}

\def\!axisinvisible{%
  \!axisvisiblefalse
  \!nextkeyword}

\def\!axislabel#1 {%
  \!axisLaBeL={#1}%
  \!placeaxislabeltrue
  \!nextkeyword}

\expandafter\def\csname !axis/\endcsname{%
  \!axissetup 
  \if!placeaxislabel
    \!placeaxislabel
  \fi
  \if +\!tickysign 
    \!dimenA=\!axisylevel
    \advance\!dimenA \!offset 
    \advance\!dimenA -\!areatloc 
    \ifdim \!dimenA>\!plotheadingoffset
      \!plotheadingoffset=\!dimenA 
    \fi
  \fi}

\def\grid #1 #2 {%
  \!countA=#1\advance\!countA 1
  \axis bottom invisible ticks length <\!zpt> andacross quantity {\!countA} /
  \!countA=#2\advance\!countA 1
  \axis left   invisible ticks length <\!zpt> andacross quantity {\!countA} / }

\def\plotheading#1 {%
  \advance\!plotheadingoffset \headingtoplotskip
  \!dimenput {#1} [B] <.5\!xaxislength,\!plotheadingoffset>
    (\!arealloc,\!areatloc)}

\def\!axisxsetup{%
  \!axisxlevel=\!arealloc
  \!axisstart=\!arealloc
  \!axisend=\!arearloc
  \!axisLength=\!xaxislength
  \!!origin=\!xorigin
  \!!unit=\!xunit
  \!xswitchtrue
  \if!axisvisible 
    \!makeaxis
  \fi}

\def\!axisysetup{%
  \!axisylevel=\!areabloc
  \!axisstart=\!areabloc
  \!axisend=\!areatloc
  \!axisLength=\!yaxislength
  \!!origin=\!yorigin
  \!!unit=\!yunit
  \!xswitchfalse
  \if!axisvisible
    \!makeaxis
  \fi}

\def\!makeaxis{%
  \setbox\!boxA=\hbox{
    \beginpicture
      \!setdimenmode
      \setcoordinatesystem point at {\!zpt} {\!zpt}   
      \putrule from {\!zpt} {\!zpt} to
        {\!tickysign\!tickysign\!axisLength} 
        {\!tickxsign\!tickxsign\!axisLength}
    \endpicturesave <\!Xsave,\!Ysave>}%
    \wd\!boxA=\!zpt
    \!placetick\!axisstart}

\def\!placeaxislabel{%
  \advance\!offset \valuestolabelleading
  \if!xswitch
    \!dimenput {\the\!axisLaBeL} [\!axislabeltbrl]
      <.5\!axisLength,\!tickysign\!offset> (\!axisxlevel,\!axisylevel)
    \advance\!offset \!dp  
    \advance\!offset \!ht  
  \else
    \!dimenput {\the\!axisLaBeL} [\!axislabeltbrl]
      <\!tickxsign\!offset,.5\!axisLength> (\!axisxlevel,\!axisylevel)
  \fi
  \!axisLaBeL={}}

%


\def\arrow <#1> [#2,#3]{%
  \!ifnextchar<{\!arrow{#1}{#2}{#3}}{\!arrow{#1}{#2}{#3}<\!zpt,\!zpt> }}

\def\!arrow#1#2#3<#4,#5> from #6 #7 to #8 #9 {%
%
  \!xloc=\!M{#8}\!xunit   
  \!yloc=\!M{#9}\!yunit
  \!dxpos=\!xloc  \!dimenA=\!M{#6}\!xunit  \advance \!dxpos -\!dimenA
  \!dypos=\!yloc  \!dimenA=\!M{#7}\!yunit  \advance \!dypos -\!dimenA
  \let\!MAH=\!M
  \!setdimenmode
  \!xshift=#4\relax  \!yshift=#5\relax
  \!reverserotateonly\!xshift\!yshift
  \advance\!xshift\!xloc  \advance\!yshift\!yloc
%
  \!xS=-\!dxpos  \advance\!xS\!xshift
  \!yS=-\!dypos  \advance\!yS\!yshift
  \!start (\!xS,\!yS)
  \!ljoin (\!xshift,\!yshift)
%
  \!Pythag\!dxpos\!dypos\!arclength
  \!divide\!dxpos\!arclength\!dxpos  
  \!dxpos=32\!dxpos  \!removept\!dxpos\!!cos
  \!divide\!dypos\!arclength\!dypos  
  \!dypos=32\!dypos  \!removept\!dypos\!!sin
%
  \!halfhead{#1}{#2}{#3}
  \!halfhead{#1}{-#2}{-#3}
  \let\!M=\!MAH
  \ignorespaces}
%
  \def\!halfhead#1#2#3{%
    \!dimenC=-#1%
    \divide \!dimenC 2 
    \!dimenD=#2\!dimenC
    \!rotate(\!dimenC,\!dimenD)by(\!!cos,\!!sin)to(\!xM,\!yM)
    \!dimenC=-#1
    \!dimenD=#3\!dimenC
    \!dimenD=.5\!dimenD
    \!rotate(\!dimenC,\!dimenD)by(\!!cos,\!!sin)to(\!xE,\!yE)
    \!start (\!xshift,\!yshift)
    \advance\!xM\!xshift  \advance\!yM\!yshift
    \advance\!xE\!xshift  \advance\!yE\!yshift
    \!qjoin (\!xM,\!yM) (\!xE,\!yE) 
    \ignorespaces}

\def\betweenarrows #1#2 from #3 #4 to #5 #6 {%
  \!xloc=\!M{#3}\!xunit  \!xxloc=\!M{#5}\!xunit%
  \!yloc=\!M{#4}\!yunit  \!yyloc=\!M{#6}\!yunit%
  \!dxpos=\!xxloc  \advance\!dxpos by -\!xloc
  \!dypos=\!yyloc  \advance\!dypos by -\!yloc
  \advance\!xloc .5\!dxpos
  \advance\!yloc .5\!dypos
  \let\!MBA=\!M
  \!setdimenmode
  \ifdim\!dypos=\!zpt
    \ifdim\!dxpos<\!zpt \!dxpos=-\!dxpos \fi
    \put {\!lrarrows{\!dxpos}{#1}}#2{} at {\!xloc} {\!yloc}
  \else
    \ifdim\!dxpos=\!zpt
      \ifdim\!dypos<\!zpt \!dypos=-\!zpt \fi
      \put {\!udarrows{\!dypos}{#1}}#2{} at {\!xloc} {\!yloc}
    \fi
  \fi
  \let\!M=\!MBA
  \ignorespaces}

\def\!lrarrows#1#2{
  {\setbox\!boxA=\hbox{$\mkern-2mu\mathord-\mkern-2mu$}%
   \setbox\!boxB=\hbox{$\leftarrow$}\!dimenE=\ht\!boxB
   \setbox\!boxB=\hbox{}\ht\!boxB=2\!dimenE
   \hbox to #1{$\mathord\leftarrow\mkern-6mu
     \cleaders\copy\!boxA\hfil
     \mkern-6mu\mathord-$%
     \kern.4em $\vcenter{\box\!boxB}$$\vcenter{\hbox{#2}}$\kern.4em
     $\mathord-\mkern-6mu
     \cleaders\copy\!boxA\hfil
     \mkern-6mu\mathord\rightarrow$}}}

\def\!udarrows#1#2{
  {\setbox\!boxB=\hbox{#2}%
   \setbox\!boxA=\hbox to \wd\!boxB{\hss$\vert$\hss}%
   \!dimenE=\ht\!boxA \advance\!dimenE \dp\!boxA \divide\!dimenE 2
   \vbox to #1{\offinterlineskip
      \vskip .05556\!dimenE
      \hbox to \wd\!boxB{\hss$\mkern.4mu\uparrow$\hss}\vskip-\!dimenE
      \cleaders\copy\!boxA\vfil
      \vskip-\!dimenE\copy\!boxA
      \vskip\!dimenE\copy\!boxB\vskip.4em
      \copy\!boxA\vskip-\!dimenE
      \cleaders\copy\!boxA\vfil
      \vskip-\!dimenE \hbox to \wd\!boxB{\hss$\mkern.4mu\downarrow$\hss}
      \vskip .05556\!dimenE}}}

%

\def\putbar#1breadth <#2> from #3 #4 to #5 #6 {%
  \!xloc=\!M{#3}\!xunit  \!xxloc=\!M{#5}\!xunit%
  \!yloc=\!M{#4}\!yunit  \!yyloc=\!M{#6}\!yunit%
  \!dypos=\!yyloc  \advance\!dypos by -\!yloc
  \!dimenI=#2  
  \ifdim \!dimenI=\!zpt 
    \putrule#1from {#3} {#4} to {#5} {#6} 
  \else 
    \let\!MBar=\!M
    \!setdimenmode 
    \divide\!dimenI 2
    \ifdim \!dypos=\!zpt             
      \advance \!yloc -\!dimenI 
      \advance \!yyloc \!dimenI
    \else
      \advance \!xloc -\!dimenI 
      \advance \!xxloc \!dimenI
    \fi
    \putrectangle#1corners at {\!xloc} {\!yloc} and {\!xxloc} {\!yyloc}
    \let\!M=\!MBar 
  \fi
  \ignorespaces}

\def\setbars#1breadth <#2> baseline at #3 = #4 {%
  \edef\!barshift{#1}%
  \edef\!barbreadth{#2}%
  \edef\!barorientation{#3}%
  \edef\!barbaseline{#4}%
  \def\!bardobaselabel{\!bardoendlabel}%
  \def\!bardoendlabel{\!barfinish}%
  \let\!drawcurve=\!barcurve
  \!setbars}
\def\!setbars{%
  \futurelet\!nextchar\!!setbars}
\def\!!setbars{%
  \if b\!nextchar
    \def\!!!setbars{\!setbarsbget}%
  \else 
    \if e\!nextchar
      \def\!!!setbars{\!setbarseget}%
    \else
      \def\!!!setbars{\relax}%
    \fi
  \fi
  \!!!setbars}
\def\!setbarsbget baselabels (#1) {%
  \def\!barbaselabelorientation{#1}%
  \def\!bardobaselabel{\!!bardobaselabel}%
  \!setbars}
\def\!setbarseget endlabels (#1) {%
  \edef\!barendlabelorientation{#1}%
  \def\!bardoendlabel{\!!bardoendlabel}%
  \!setbars}

\def\!barcurve #1 #2 {%
  \if y\!barorientation
    \def\!basexarg{#1}%
    \def\!baseyarg{\!barbaseline}%
  \else
    \def\!basexarg{\!barbaseline}%
    \def\!baseyarg{#2}%
  \fi
  \expandafter\putbar\!barshift breadth <\!barbreadth> from {\!basexarg}
    {\!baseyarg} to {#1} {#2}
  \def\!endxarg{#1}%
  \def\!endyarg{#2}%
  \!bardobaselabel}

\def\!!bardobaselabel "#1" {%
  \put {#1}\!barbaselabelorientation{} at {\!basexarg} {\!baseyarg}
  \!bardoendlabel}

\def\!!bardoendlabel "#1" {%
  \put {#1}\!barendlabelorientation{} at {\!endxarg} {\!endyarg}
  \!barfinish}

\def\!barfinish{%
  \!ifnextchar/{\!finish}{\!barcurve}}

%
%
%
\def\putrectangle{%
  \!ifnextchar<{\!putrectangle}{\!putrectangle<\!zpt,\!zpt> }}
\def\!putrectangle<#1,#2> corners at #3 #4 and #5 #6 {%
%
  \!xone=\!M{#3}\!xunit  \!xtwo=\!M{#5}\!xunit%
  \!yone=\!M{#4}\!yunit  \!ytwo=\!M{#6}\!yunit%
  \ifdim \!xtwo<\!xone
    \!dimenI=\!xone  \!xone=\!xtwo  \!xtwo=\!dimenI
  \fi
  \ifdim \!ytwo<\!yone
    \!dimenI=\!yone  \!yone=\!ytwo  \!ytwo=\!dimenI
  \fi
  \!dimenI=#1\relax  \advance\!xone\!dimenI  \advance\!xtwo\!dimenI
  \!dimenI=#2\relax  \advance\!yone\!dimenI  \advance\!ytwo\!dimenI
  \let\!MRect=\!M
  \!setdimenmode
%
  \!shaderectangle
%
  \!dimenI=.5\linethickness
  \advance \!xone  -\!dimenI
  \advance \!xtwo   \!dimenI
  \putrule from {\!xone} {\!yone} to {\!xtwo} {\!yone} 
  \putrule from {\!xone} {\!ytwo} to {\!xtwo} {\!ytwo} 
%
  \advance \!xone   \!dimenI
  \advance \!xtwo  -\!dimenI%
  \advance \!yone  -\!dimenI
  \advance \!ytwo   \!dimenI
  \putrule from {\!xone} {\!yone} to {\!xone} {\!ytwo} 
  \putrule from {\!xtwo} {\!yone} to {\!xtwo} {\!ytwo} 
  \let\!M=\!MRect
  \ignorespaces}

\def\shaderectangleson{%
  \def\!shaderectangle{\!!shaderectangle}%
  \ignorespaces}
\def\shaderectanglesoff{%
  \def\!shaderectangle{}%
  \ignorespaces}

\shaderectanglesoff

\def\!!shaderectangle{%
  \!dimenA=\!xtwo  \advance \!dimenA -\!xone
  \!dimenB=\!ytwo  \advance \!dimenB -\!yone
  \ifdim \!dimenA<\!dimenB
    \!startvshade (\!xone,\!yone,\!ytwo)
    \!lshade      (\!xtwo,\!yone,\!ytwo)
  \else
    \!starthshade (\!yone,\!xone,\!xtwo)
    \!lshade      (\!ytwo,\!xone,\!xtwo)
  \fi
  \ignorespaces}
  
\def\frame{%
  \!ifnextchar<{\!frame}{\!frame<\!zpt> }}
\long\def\!frame<#1> #2{%
  \beginpicture
    \setcoordinatesystem units <1pt,1pt> point at 0 0 
    \put {#2} [Bl] at 0 0 
    \!dimenA=#1\relax
    \!dimenB=\!wd \advance \!dimenB \!dimenA
    \!dimenC=\!ht \advance \!dimenC \!dimenA
    \!dimenD=\!dp \advance \!dimenD \!dimenA
    \let\!MFr=\!M
    \!setdimenmode
    \putrectangle corners at {-\!dimenA} {-\!dimenD} and {\!dimenB} {\!dimenC}
    \!setcoordmode
    \let\!M=\!MFr
  \endpicture
  \ignorespaces}

\def\rectangle <#1> <#2> {%
  \setbox0=\hbox{}\wd0=#1\ht0=#2\frame {\box0}}

%

\def\plot{%
  \!ifnextchar"{\!plotfromfile}{\!drawcurve}}
\def\!plotfromfile"#1"{%
  \expandafter\!drawcurve \input #1 /}

\def\setquadratic{%
  \let\!drawcurve=\!qcurve
  \let\!!Shade=\!!qShade
  \let\!!!Shade=\!!!qShade}

\def\setlinear{%
  \let\!drawcurve=\!lcurve
  \let\!!Shade=\!!lShade
  \let\!!!Shade=\!!!lShade}

\def\sethistograms{%
  \let\!drawcurve=\!hcurve}

\def\!qcurve #1 #2 {%
  \!start (#1,#2)
  \!Qjoin}
\def\!Qjoin#1 #2 #3 #4 {%
  \!qjoin (#1,#2) (#3,#4)             
  \!ifnextchar/{\!finish}{\!Qjoin}}

\def\!lcurve #1 #2 {%
  \!start (#1,#2)
  \!Ljoin}
\def\!Ljoin#1 #2 {%
  \!ljoin (#1,#2)                    
  \!ifnextchar/{\!finish}{\!Ljoin}}

\def\!finish/{\ignorespaces}

\def\!hcurve #1 #2 {%
  \edef\!hxS{#1}%
  \edef\!hyS{#2}%
  \!hjoin}
\def\!hjoin#1 #2 {%
  \putrectangle corners at {\!hxS} {\!hyS} and {#1} {#2}
  \edef\!hxS{#1}%
  \!ifnextchar/{\!finish}{\!hjoin}}

\def\vshade #1 #2 #3 {%
  \!startvshade (#1,#2,#3)
  \!Shadewhat}

\def\hshade #1 #2 #3 {%
  \!starthshade (#1,#2,#3)
  \!Shadewhat}

\def\!Shadewhat{%
  \futurelet\!nextchar\!Shade}
\def\!Shade{%
  \if <\!nextchar
    \def\!nextShade{\!!Shade}%
  \else
    \if /\!nextchar
      \def\!nextShade{\!finish}%
    \else
      \def\!nextShade{\!!!Shade}%
    \fi
  \fi
  \!nextShade}
\def\!!lShade<#1> #2 #3 #4 {%
  \!lshade <#1> (#2,#3,#4)                 
  \!Shadewhat}
\def\!!!lShade#1 #2 #3 {%
  \!lshade (#1,#2,#3)
  \!Shadewhat} 
\def\!!qShade<#1> #2 #3 #4 #5 #6 #7 {%
  \!qshade <#1> (#2,#3,#4) (#5,#6,#7)      
  \!Shadewhat}
\def\!!!qShade#1 #2 #3 #4 #5 #6 {%
  \!qshade (#1,#2,#3) (#4,#5,#6)
  \!Shadewhat} 

\setlinear

\def\setdashpattern <#1>{%
  \def\!Flist{}\def\!Blist{}\def\!UDlist{}%
  \!countA=0
  \!ecfor\!item:=#1\do{%
    \!dimenA=\!item\relax
    \expandafter\!rightappend\the\!dimenA\withCS{\\}\to\!UDlist%
    \advance\!countA  1
    \ifodd\!countA
      \expandafter\!rightappend\the\!dimenA\withCS{\!Rule}\to\!Flist%
      \expandafter\!leftappend\the\!dimenA\withCS{\!Rule}\to\!Blist%
    \else 
      \expandafter\!rightappend\the\!dimenA\withCS{\!Skip}\to\!Flist%
      \expandafter\!leftappend\the\!dimenA\withCS{\!Skip}\to\!Blist%
    \fi}%
  \!leaderlength=\!zpt
  \def\!Rule##1{\advance\!leaderlength  ##1}%
  \def\!Skip##1{\advance\!leaderlength  ##1}%
  \!Flist%
  \ifdim\!leaderlength>\!zpt 
  \else
    \def\!Flist{\!Skip{24in}}\def\!Blist{\!Skip{24in}}\ignorespaces
    \def\!UDlist{\\{\!zpt}\\{24in}}\ignorespaces
    \!leaderlength=24in
  \fi
  \!dashingon}

\def\!dashingon{%
  \def\!advancedashing{\!!advancedashing}%
  \def\!drawlinearsegment{\!lineardashed}%
  \def\!puthline{\!putdashedhline}%
  \def\!putvline{\!putdashedvline}%
  \ignorespaces}%
\def\!dashingoff{%
  \def\!advancedashing{\relax}%
  \def\!drawlinearsegment{\!linearsolid}%
  \def\!puthline{\!putsolidhline}%
  \def\!putvline{\!putsolidvline}%
  \ignorespaces}

\def\setdots{%
  \!ifnextchar<{\!setdots}{\!setdots<5pt>}}
\def\!setdots<#1>{%
  \!dimenB=#1\advance\!dimenB -\plotsymbolspacing
  \ifdim\!dimenB<\!zpt
    \!dimenB=\!zpt
  \fi
\setdashpattern <\plotsymbolspacing,\!dimenB>}

\def\setdotsnear <#1> for <#2>{%
  \!dimenB=#2\relax  \advance\!dimenB -.05pt  
  \!dimenC=#1\relax  \!countA=\!dimenC 
  \!dimenD=\!dimenB  \advance\!dimenD .5\!dimenC  \!countB=\!dimenD
  \divide \!countB  \!countA
  \ifnum 1>\!countB 
    \!countB=1
  \fi
  \divide\!dimenB  \!countB
  \setdots <\!dimenB>}

\def\setdashes{%
  \!ifnextchar<{\!setdashes}{\!setdashes<5pt>}}
\def\!setdashes<#1>{\setdashpattern <#1,#1>}

\def\setdashesnear <#1> for <#2>{%
  \!dimenB=#2\relax  
  \!dimenC=#1\relax  \!countA=\!dimenC 
  \!dimenD=\!dimenB  \advance\!dimenD .5\!dimenC  \!countB=\!dimenD
  \divide \!countB  \!countA
  \ifodd \!countB 
  \else 
    \advance \!countB  1
  \fi
  \divide\!dimenB  \!countB
  \setdashes <\!dimenB>}

\def\setsolid{%
  \def\!Flist{\!Rule{24in}}\def\!Blist{\!Rule{24in}}%
  \def\!UDlist{\\{24in}\\{\!zpt}}%
  \!dashingoff}  
\setsolid

\def\findlength#1{%
  \begingroup
    \setdashpattern <0pt, \maxdimen>
    \setplotsymbol ({})  
    \dontsavelinesandcurves
    #1%
  \endgroup
  \ignorespaces}


  

\def\!divide#1#2#3{%
  \!dimenB=#1
  \!dimenC=#2
  \!dimenD=\!dimenB
  \divide \!dimenD \!dimenC
  \!dimenA=\!dimenD
  \multiply\!dimenD \!dimenC
  \advance\!dimenB -\!dimenD
  \!dimenD=\!dimenC
    \ifdim\!dimenD<\!zpt \!dimenD=-\!dimenD 
  \fi
  \ifdim\!dimenD<64pt
    \!divstep[\!tfs]\!divstep[\!tfs]%
  \else 
    \!!divide
  \fi
  #3=\!dimenA\ignorespaces}

\def\!!divide{%
  \ifdim\!dimenD<256pt
    \!divstep[64]\!divstep[32]\!divstep[32]%
  \else 
    \!divstep[8]\!divstep[8]\!divstep[8]\!divstep[8]\!divstep[8]%
    \!dimenA=2\!dimenA
  \fi}

\def\!divstep[#1]{
  \!dimenB=#1\!dimenB
  \!dimenD=\!dimenB
    \divide \!dimenD by \!dimenC
  \!dimenA=#1\!dimenA
    \advance\!dimenA by \!dimenD%
  \multiply\!dimenD by \!dimenC
    \advance\!dimenB by -\!dimenD}

\def\Divide <#1> by <#2> forming <#3> {%
  \!divide{#1}{#2}{#3}}





\def\circulararc{%
  \ellipticalarc axes ratio 1:1 }

\def\ellipticalarc axes ratio #1:#2 #3 degrees from #4 #5 center at #6 #7 {%
  \!angle=#3pt\relax
  \ifdim\!angle>\!zpt 
    \def\!sign{}
  \else 
    \def\!sign{-}\!angle=-\!angle
  \fi
  \!xxloc=\!M{#6}\!xunit
  \!yyloc=\!M{#7}\!yunit     
  \!xxS=\!M{#4}\!xunit
  \!yyS=\!M{#5}\!yunit
  \advance\!xxS -\!xxloc
  \advance\!yyS -\!yyloc
  \!divide\!xxS{#1pt}\!xxS 
  \!divide\!yyS{#2pt}\!yyS 
  \let\!MC=\!M
  \!setdimenmode
  \!xS=#1\!xxS  \advance\!xS\!xxloc
  \!yS=#2\!yyS  \advance\!yS\!yyloc
  \!start (\!xS,\!yS)%
  \!loop\ifdim\!angle>14.9999pt
    \!rotate(\!xxS,\!yyS)by(\!cos,\!sign\!sin)to(\!xxM,\!yyM) 
    \!rotate(\!xxM,\!yyM)by(\!cos,\!sign\!sin)to(\!xxE,\!yyE)
    \!xM=#1\!xxM  \advance\!xM\!xxloc  \!yM=#2\!yyM  \advance\!yM\!yyloc
    \!xE=#1\!xxE  \advance\!xE\!xxloc  \!yE=#2\!yyE  \advance\!yE\!yyloc
    \!qjoin (\!xM,\!yM) (\!xE,\!yE)
    \!xxS=\!xxE  \!yyS=\!yyE 
    \advance \!angle -15pt
  \repeat
  \ifdim\!angle>\!zpt
    \!angle=100.53096\!angle
    \divide \!angle 360 
    \!sinandcos\!angle\!!sin\!!cos
    \!rotate(\!xxS,\!yyS)by(\!!cos,\!sign\!!sin)to(\!xxM,\!yyM) 
    \!rotate(\!xxM,\!yyM)by(\!!cos,\!sign\!!sin)to(\!xxE,\!yyE)
    \!xM=#1\!xxM  \advance\!xM\!xxloc  \!yM=#2\!yyM  \advance\!yM\!yyloc
    \!xE=#1\!xxE  \advance\!xE\!xxloc  \!yE=#2\!yyE  \advance\!yE\!yyloc
    \!qjoin (\!xM,\!yM) (\!xE,\!yE)
  \fi
  \let\!M=\!MC
  \ignorespaces}

\def\!rotate(#1,#2)by(#3,#4)to(#5,#6){%
  \!dimenA=#3#1\advance \!dimenA -#4#2
  \!dimenB=#3#2\advance \!dimenB  #4#1
  \divide \!dimenA 32  \divide \!dimenB 32 
  #5=\!dimenA  #6=\!dimenB
  \ignorespaces}
\def\!sin{4.17684}
\def\!cos{31.72624}

\def\!sinandcos#1#2#3{%
 \!dimenD=#1
 \!dimenA=\!dimenD
 \!dimenB=32pt
 \!removept\!dimenD\!value
 \!dimenC=\!dimenD
 \!dimenC=\!value\!dimenC \divide\!dimenC by 64 
 \advance\!dimenB by -\!dimenC
 \!dimenC=\!value\!dimenC \divide\!dimenC by 96 
 \advance\!dimenA by -\!dimenC
 \!dimenC=\!value\!dimenC \divide\!dimenC by 128 
 \advance\!dimenB by \!dimenC%
 \!removept\!dimenA#2
 \!removept\!dimenB#3
 \ignorespaces}




\def\putrule#1from #2 #3 to #4 #5 {%
  \!xloc=\!M{#2}\!xunit  \!xxloc=\!M{#4}\!xunit%
  \!yloc=\!M{#3}\!yunit  \!yyloc=\!M{#5}\!yunit%
  \!dxpos=\!xxloc  \advance\!dxpos by -\!xloc
  \!dypos=\!yyloc  \advance\!dypos by -\!yloc
  \ifdim\!dypos=\!zpt
    \def\!!Line{\!puthline{#1}}\ignorespaces
  \else
    \ifdim\!dxpos=\!zpt
      \def\!!Line{\!putvline{#1}}\ignorespaces
    \else 
       \def\!!Line{}
    \fi
  \fi
  \let\!ML=\!M
  \!setdimenmode
  \!!Line%
  \let\!M=\!ML
  \ignorespaces}

\def\!putsolidhline#1{%
  \ifdim\!dxpos>\!zpt 
    \put{\!hline\!dxpos}#1[l] at {\!xloc} {\!yloc}
  \else 
    \put{\!hline{-\!dxpos}}#1[l] at {\!xxloc} {\!yyloc}
  \fi
  \ignorespaces}

\def\!putsolidvline#1{%
  \ifdim\!dypos>\!zpt 
    \put{\!vline\!dypos}#1[b] at {\!xloc} {\!yloc}
  \else 
    \put{\!vline{-\!dypos}}#1[b] at {\!xxloc} {\!yyloc}
  \fi
  \ignorespaces}

\def\!hline#1{\hbox to #1{\leaders \hrule height\linethickness\hfill}}
\def\!vline#1{\vbox to #1{\leaders \vrule width\linethickness\vfill}}

\def\!putdashedhline#1{%
  \ifdim\!dxpos>\!zpt 
    \!DLsetup\!Flist\!dxpos
    \put{\hbox to \!totalleaderlength{\!hleaders}\!hpartialpattern\!Rtrunc}
      #1[l] at {\!xloc} {\!yloc} 
  \else 
    \!DLsetup\!Blist{-\!dxpos}
    \put{\!hpartialpattern\!Ltrunc\hbox to \!totalleaderlength{\!hleaders}}
      #1[r] at {\!xloc} {\!yloc} 
  \fi
  \ignorespaces}

\def\!putdashedvline#1{%
  \!dypos=-\!dypos
  \ifdim\!dypos>\!zpt 
    \!DLsetup\!Flist\!dypos 
    \put{\vbox{\vbox to \!totalleaderlength{\!vleaders}
      \!vpartialpattern\!Rtrunc}}#1[t] at {\!xloc} {\!yloc} 
  \else 
    \!DLsetup\!Blist{-\!dypos}
    \put{\vbox{\!vpartialpattern\!Ltrunc
      \vbox to \!totalleaderlength{\!vleaders}}}#1[b] at {\!xloc} {\!yloc} 
  \fi
  \ignorespaces}

\def\!DLsetup#1#2{
  \let\!RSlist=#1
  \!countB=#2
  \!countA=\!leaderlength
  \divide\!countB by \!countA
  \!totalleaderlength=\!countB\!leaderlength
  \!Rresiduallength=#2%
  \advance \!Rresiduallength by -\!totalleaderlength
  \!Lresiduallength=\!leaderlength
  \advance \!Lresiduallength by -\!Rresiduallength
  \ignorespaces}

\def\!hleaders{%
  \def\!Rule##1{\vrule height\linethickness width##1}%
  \def\!Skip##1{\hskip##1}%
  \leaders\hbox{\!RSlist}\hfill}

\def\!hpartialpattern#1{%
  \!dimenA=\!zpt \!dimenB=\!zpt 
  \def\!Rule##1{#1{##1}\vrule height\linethickness width\!dimenD}%
  \def\!Skip##1{#1{##1}\hskip\!dimenD}%
  \!RSlist}

\def\!vleaders{%
  \def\!Rule##1{\hrule width\linethickness height##1}%
  \def\!Skip##1{\vskip##1}%
  \leaders\vbox{\!RSlist}\vfill}

\def\!vpartialpattern#1{%
  \!dimenA=\!zpt \!dimenB=\!zpt 
  \def\!Rule##1{#1{##1}\hrule width\linethickness height\!dimenD}%
  \def\!Skip##1{#1{##1}\vskip\!dimenD}%
  \!RSlist}

\def\!Rtrunc#1{\!trunc{#1}>\!Rresiduallength}
\def\!Ltrunc#1{\!trunc{#1}<\!Lresiduallength}

\def\!trunc#1#2#3{%
  \!dimenA=\!dimenB         
  \advance\!dimenB by #1%
  \!dimenD=\!dimenB  \ifdim\!dimenD#2#3\!dimenD=#3\fi
  \!dimenC=\!dimenA  \ifdim\!dimenC#2#3\!dimenC=#3\fi
  \advance \!dimenD by -\!dimenC}

\def\!start (#1,#2){%
  \!plotxorigin=\!xorigin  \advance \!plotxorigin by \!plotsymbolxshift
  \!plotyorigin=\!yorigin  \advance \!plotyorigin by \!plotsymbolyshift
  \!xS=\!M{#1}\!xunit \!yS=\!M{#2}\!yunit
  \!rotateaboutpivot\!xS\!yS
  \!copylist\!UDlist\to\!!UDlist
  \!getnextvalueof\!downlength\from\!!UDlist
  \!distacross=\!zpt
  \!intervalno=0 
  \global\totalarclength=\!zpt
  \ignorespaces}

\def\!ljoin (#1,#2){%
  \advance\!intervalno by 1
  \!xE=\!M{#1}\!xunit \!yE=\!M{#2}\!yunit
  \!rotateaboutpivot\!xE\!yE
  \!xdiff=\!xE \advance \!xdiff by -\!xS
  \!ydiff=\!yE \advance \!ydiff by -\!yS
  \!Pythag\!xdiff\!ydiff\!arclength
  \global\advance \totalarclength by \!arclength%
  \!drawlinearsegment
  \!xS=\!xE \!yS=\!yE
  \ignorespaces}

\def\!linearsolid{%
  \!npoints=\!arclength
  \!countA=\plotsymbolspacing
  \divide\!npoints by \!countA
  \ifnum \!npoints<1 
    \!npoints=1 
  \fi
  \divide\!xdiff by \!npoints
  \divide\!ydiff by \!npoints
  \!xpos=\!xS \!ypos=\!yS
  \loop\ifnum\!npoints>-1
    \!plotifinbounds
    \advance \!xpos by \!xdiff
    \advance \!ypos by \!ydiff
    \advance \!npoints by -1
  \repeat
  \ignorespaces}

\def\!lineardashed{%
  \ifdim\!distacross>\!arclength
    \advance \!distacross by -\!arclength  
  \else
    \loop\ifdim\!distacross<\!arclength
      \!divide\!distacross\!arclength\!dimenA
      \!removept\!dimenA\!t
      \!xpos=\!t\!xdiff \advance \!xpos by \!xS
      \!ypos=\!t\!ydiff \advance \!ypos by \!yS
      \!plotifinbounds
      \advance\!distacross by \plotsymbolspacing
      \!advancedashing
    \repeat  
    \advance \!distacross by -\!arclength
  \fi
  \ignorespaces}

\def\!!advancedashing{%
  \advance\!downlength by -\plotsymbolspacing
  \ifdim \!downlength>\!zpt
  \else
    \advance\!distacross by \!downlength
    \!getnextvalueof\!uplength\from\!!UDlist
    \advance\!distacross by \!uplength
    \!getnextvalueof\!downlength\from\!!UDlist
  \fi}

\def\inboundscheckoff{%
  \def\!plotifinbounds{\!plot(\!xpos,\!ypos)}%
  \def\!initinboundscheck{\relax}\ignorespaces}
\def\inboundscheckon{%
  \def\!plotifinbounds{\!!plotifinbounds}%
  \def\!initinboundscheck{\!!initinboundscheck}%
  \!initinboundscheck\ignorespaces} 
\inboundscheckoff

\def\!!plotifinbounds{%
  \ifdim \!xpos<\!checkleft
  \else
    \ifdim \!xpos>\!checkright
    \else
      \ifdim \!ypos<\!checkbot
      \else
         \ifdim \!ypos>\!checktop
         \else
           \!plot(\!xpos,\!ypos)
         \fi 
      \fi
    \fi
  \fi}

\def\!!initinboundscheck{%
  \!checkleft=\!arealloc     \advance\!checkleft by \!xorigin
  \!checkright=\!arearloc    \advance\!checkright by \!xorigin
  \!checkbot=\!areabloc      \advance\!checkbot by \!yorigin
  \!checktop=\!areatloc      \advance\!checktop by \!yorigin}

%


\def\!logten#1#2{%
  \expandafter\!!logten#1\!nil
  \!removept\!dimenF#2%
  \ignorespaces}

\def\!!logten#1#2\!nil{%
  \if -#1%
    \!dimenF=\!zpt
    \def\!next{\ignorespaces}%
  \else
    \if +#1%
      \def\!next{\!!logten#2\!nil}%
    \else
      \if .#1%
        \def\!next{\!!logten0.#2\!nil}%
      \else
        \def\!next{\!!!logten#1#2..\!nil}%
      \fi
    \fi
  \fi
  \!next}

\def\!!!logten#1#2.#3.#4\!nil{%
  \!dimenF=1pt 
  \if 0#1%
    \!!logshift#3pt 
  \else 
    \!logshift#2/
    \!dimenE=#1.#2#3pt 
  \fi 
  \ifdim \!dimenE<\!rootten
    \multiply \!dimenE 10 
    \advance  \!dimenF -1pt
  \fi
  \!dimenG=\!dimenE
    \advance\!dimenG 10pt
  \advance\!dimenE -10pt 
  \multiply\!dimenE 10 
  \!divide\!dimenE\!dimenG\!dimenE
  \!removept\!dimenE\!t
  \!dimenG=\!t\!dimenE
  \!removept\!dimenG\!tt
  \!dimenH=\!tt\!tenAe
    \divide\!dimenH 100
  \advance\!dimenH \!tenAc
  \!dimenH=\!tt\!dimenH
    \divide\!dimenH 100   
  \advance\!dimenH \!tenAa
  \!dimenH=\!t\!dimenH
    \divide\!dimenH 100 
  \advance\!dimenF \!dimenH}

\def\!logshift#1{%
  \if #1/%
    \def\!next{\ignorespaces}%
  \else
    \advance\!dimenF 1pt 
    \def\!next{\!logshift}%
  \fi 
  \!next}

 \def\!!logshift#1{%
   \advance\!dimenF -1pt
   \if 0#1%
     \def\!next{\!!logshift}%
   \else
     \if p#1%
       \!dimenF=1pt
       \def\!next{\!dimenE=1p}%
     \else
       \def\!next{\!dimenE=#1.}%
     \fi
   \fi
   \!next}

\def\beginpicture{%
  \setbox\!picbox=\hbox\bgroup%
  \!xleft=\maxdimen  
  \!xright=-\maxdimen
  \!ybot=\maxdimen
  \!ytop=-\maxdimen}

\def\endpicture{%
  \ifdim\!xleft=\maxdimen
    \!xleft=\!zpt \!xright=\!zpt \!ybot=\!zpt \!ytop=\!zpt 
  \fi
  \global\!Xleft=\!xleft \global\!Xright=\!xright
  \global\!Ybot=\!ybot \global\!Ytop=\!ytop
  \egroup%
  \ht\!picbox=\!Ytop  \dp\!picbox=-\!Ybot
  \ifdim\!Ybot>\!zpt
  \else 
    \ifdim\!Ytop<\!zpt
      \!Ybot=\!Ytop
    \else
      \!Ybot=\!zpt
    \fi
  \fi
  \hbox{\kern-\!Xleft\lower\!Ybot\box\!picbox\kern\!Xright}}

\def\endpicturesave <#1,#2>{%
  \endpicture \global #1=\!Xleft \global #2=\!Ybot \ignorespaces}

\def\setcoordinatesystem{%
  \!ifnextchar{u}{\!getlengths }
    {\!getlengths units <\!xunit,\!yunit>}}
\def\!getlengths units <#1,#2>{%
  \!xunit=#1\relax
  \!yunit=#2\relax
  \!ifcoordmode 
    \let\!SCnext=\!SCccheckforRP
  \else
    \let\!SCnext=\!SCdcheckforRP
  \fi
  \!SCnext}
\def\!SCccheckforRP{%
  \!ifnextchar{p}{\!cgetreference }
    {\!cgetreference point at {\!xref} {\!yref} }}
\def\!cgetreference point at #1 #2 {%
  \edef\!xref{#1}\edef\!yref{#2}%
  \!xorigin=\!xref\!xunit  \!yorigin=\!yref\!yunit  
  \!initinboundscheck 
  \ignorespaces}
\def\!SCdcheckforRP{%
  \!ifnextchar{p}{\!dgetreference}%
    {\ignorespaces}}
\def\!dgetreference point at #1 #2 {%
  \!xorigin=#1\relax  \!yorigin=#2\relax
  \ignorespaces}

\long\def\put#1#2 at #3 #4 {%
  \!setputobject{#1}{#2}%
  \!xpos=\!M{#3}\!xunit  \!ypos=\!M{#4}\!yunit  
  \!rotateaboutpivot\!xpos\!ypos%
  \advance\!xpos -\!xorigin  \advance\!xpos -\!xshift
  \advance\!ypos -\!yorigin  \advance\!ypos -\!yshift
  \kern\!xpos\raise\!ypos\box\!putobject\kern-\!xpos%
  \!doaccounting\ignorespaces}

\long\def\multiput #1#2 at {%
  \!setputobject{#1}{#2}%
  \!ifnextchar"{\!putfromfile}{\!multiput}}
\def\!putfromfile"#1"{%
  \expandafter\!multiput \input #1 /}
\def\!multiput{%
  \futurelet\!nextchar\!!multiput}
\def\!!multiput{%
  \if *\!nextchar
    \def\!nextput{\!alsoby}%
  \else
    \if /\!nextchar
      \def\!nextput{\!finishmultiput}%
    \else
      \def\!nextput{\!alsoat}%
    \fi
  \fi
  \!nextput}
\def\!finishmultiput/{%
  \setbox\!putobject=\hbox{}%
  \ignorespaces}

\def\!alsoat#1 #2 {%
  \!xpos=\!M{#1}\!xunit  \!ypos=\!M{#2}\!yunit  
  \!rotateaboutpivot\!xpos\!ypos%
  \advance\!xpos -\!xorigin  \advance\!xpos -\!xshift
  \advance\!ypos -\!yorigin  \advance\!ypos -\!yshift
  \kern\!xpos\raise\!ypos\copy\!putobject\kern-\!xpos%
  \!doaccounting
  \!multiput}

\def\!alsoby*#1 #2 #3 {%
  \!dxpos=\!M{#2}\!xunit \!dypos=\!M{#3}\!yunit 
  \!rotateonly\!dxpos\!dypos
  \!ntemp=#1%
  \!!loop\ifnum\!ntemp>0
    \advance\!xpos by \!dxpos  \advance\!ypos by \!dypos
    \kern\!xpos\raise\!ypos\copy\!putobject\kern-\!xpos%
    \advance\!ntemp by -1
  \repeat
  \!doaccounting 
  \!multiput}

\def\accountingon{\def\!doaccounting{\!!doaccounting}\ignorespaces}
\def\accountingoff{\def\!doaccounting{}\ignorespaces}
\accountingon
\def\!!doaccounting{%
  \!xtemp=\!xpos  
  \!ytemp=\!ypos
  \ifdim\!xtemp<\!xleft 
     \!xleft=\!xtemp 
  \fi
  \advance\!xtemp by  \!wd 
  \ifdim\!xright<\!xtemp 
    \!xright=\!xtemp
  \fi
  \advance\!ytemp by -\!dp
  \ifdim\!ytemp<\!ybot  
    \!ybot=\!ytemp
  \fi
  \advance\!ytemp by  \!dp
  \advance\!ytemp by  \!ht 
  \ifdim\!ytemp>\!ytop  
    \!ytop=\!ytemp  
  \fi}

\long\def\!setputobject#1#2{%
  \setbox\!putobject=\hbox{#1}%
  \!ht=\ht\!putobject  \!dp=\dp\!putobject  \!wd=\wd\!putobject
  \wd\!putobject=\!zpt
  \!xshift=.5\!wd   \!yshift=.5\!ht   \advance\!yshift by -.5\!dp
  \edef\!putorientation{#2}%
  \expandafter\!SPOreadA\!putorientation[]\!nil%
  \expandafter\!SPOreadB\!putorientation<\!zpt,\!zpt>\!nil\ignorespaces}

\def\!SPOreadA#1[#2]#3\!nil{\!etfor\!orientation:=#2\do\!SPOreviseshift}

\def\!SPOreadB#1<#2,#3>#4\!nil{\advance\!xshift by -#2\advance\!yshift by -#3}

\def\!SPOreviseshift{%
  \if l\!orientation 
    \!xshift=\!zpt
  \else 
    \if r\!orientation 
      \!xshift=\!wd
    \else 
      \if b\!orientation
        \!yshift=-\!dp
      \else 
        \if B\!orientation 
          \!yshift=\!zpt
        \else 
          \if t\!orientation 
            \!yshift=\!ht
          \fi 
        \fi
      \fi
    \fi
  \fi}

\long\def\!dimenput#1#2(#3,#4){%
  \!setputobject{#1}{#2}%
  \!xpos=#3\advance\!xpos by -\!xshift
  \!ypos=#4\advance\!ypos by -\!yshift
  \kern\!xpos\raise\!ypos\box\!putobject\kern-\!xpos%
  \!doaccounting\ignorespaces}

\def\!setdimenmode{%
  \let\!M=\!M!!\ignorespaces}
\def\!setcoordmode{%
  \let\!M=\!M!\ignorespaces}
\def\!ifcoordmode{%
  \ifx \!M \!M!}
\def\!ifdimenmode{%
  \ifx \!M \!M!!}
\def\!M!#1#2{#1#2} 
\def\!M!!#1#2{#1}
\!setcoordmode
\let\setdimensionmode=\!setdimenmode
\let\setcoordinatemode=\!setcoordmode

\def\Xdistance#1{%
  \!M{#1}\!xunit
  \ignorespaces}
\def\Ydistance#1{%
  \!M{#1}\!yunit
  \ignorespaces}


\def\stack{%
  \!ifnextchar[{\!stack}{\!stack[c]}}
\def\!stack[#1]{%
  \let\!lglue=\hfill \let\!rglue=\hfill
  \expandafter\let\csname !#1glue\endcsname=\relax
  \!ifnextchar<{\!!stack}{\!!stack<\stackleading>}}
\def\!!stack<#1>#2{%
  \vbox{\def\!valueslist{}\!ecfor\!value:=#2\do{%
    \expandafter\!rightappend\!value\withCS{\\}\to\!valueslist}%
    \!lop\!valueslist\to\!value
    \let\\=\cr\lineskiplimit=\maxdimen\lineskip=#1%
    \baselineskip=-1000pt\halign{\!lglue##\!rglue\cr \!value\!valueslist\cr}}%
  \ignorespaces}

\def\lines{%
  \!ifnextchar[{\!lines}{\!lines[c]}}
\def\!lines[#1]#2{%
  \let\!lglue=\hfill \let\!rglue=\hfill
  \expandafter\let\csname !#1glue\endcsname=\relax
  \vbox{\halign{\!lglue##\!rglue\cr #2\crcr}}%
  \ignorespaces}

\def\Lines{%
  \!ifnextchar[{\!Lines}{\!Lines[c]}}
\def\!Lines[#1]#2{%
  \let\!lglue=\hfill \let\!rglue=\hfill
  \expandafter\let\csname !#1glue\endcsname=\relax
  \vtop{\halign{\!lglue##\!rglue\cr #2\crcr}}%
  \ignorespaces}




\def\setplotsymbol(#1#2){%
  \!setputobject{#1}{#2}
  \setbox\!plotsymbol=\box\!putobject%
  \!plotsymbolxshift=\!xshift 
  \!plotsymbolyshift=\!yshift 
  \ignorespaces}



\def\!!plot(#1,#2){%
  \!dimenA=-\!plotxorigin \advance \!dimenA by #1
  \!dimenB=-\!plotyorigin \advance \!dimenB by #2
  \kern\!dimenA\raise\!dimenB\copy\!plotsymbol\kern-\!dimenA%
  \ignorespaces}

\def\!!!plot(#1,#2){%
  \!dimenA=-\!plotxorigin \advance \!dimenA by #1
  \!dimenB=-\!plotyorigin \advance \!dimenB by #2
  \kern\!dimenA\raise\!dimenB\copy\!plotsymbol\kern-\!dimenA%
  \!countE=\!dimenA
  \!countF=\!dimenB
  \immediate\write\!replotfile{\the\!countE,\the\!countF.}%
  \ignorespaces}

\def\savelinesandcurves on "#1" {%
  \immediate\closeout\!replotfile
  \immediate\openout\!replotfile=#1%
  \let\!plot=\!!!plot}

\def\dontsavelinesandcurves {%
  \let\!plot=\!!plot}
\dontsavelinesandcurves

{\catcode`\%=11\xdef\!Commentsignal{
\def\writesavefile#1 {%
  \immediate\write\!replotfile{\!Commentsignal #1}%
  \ignorespaces}

\def\replot"#1" {%
  \expandafter\!replot\input #1 /}
\def\!replot#1,#2. {%
  \!dimenA=#1sp
  \kern\!dimenA\raise#2sp\copy\!plotsymbol\kern-\!dimenA
  \futurelet\!nextchar\!!replot}
\def\!!replot{%
  \if /\!nextchar 
    \def\!next{\!finish}%
  \else
    \def\!next{\!replot}%
  \fi
  \!next}




\def\!Pythag#1#2#3{%
  \!dimenE=#1\relax                                     
  \ifdim\!dimenE<\!zpt 
    \!dimenE=-\!dimenE 
  \fi
  \!dimenF=#2\relax
  \ifdim\!dimenF<\!zpt 
    \!dimenF=-\!dimenF 
  \fi
  \advance \!dimenF by \!dimenE
  \ifdim\!dimenF=\!zpt 
    \!dimenG=\!zpt
  \else 
    \!divide{8\!dimenE}\!dimenF\!dimenE
    \advance\!dimenE by -4pt
      \!dimenE=2\!dimenE
    \!removept\!dimenE\!!t
    \!dimenE=\!!t\!dimenE
    \advance\!dimenE by 64pt
    \divide \!dimenE by 2
    \!dimenH=7pt
    \!!Pythag\!!Pythag\!!Pythag
    \!removept\!dimenH\!!t
    \!dimenG=\!!t\!dimenF
    \divide\!dimenG by 8
  \fi
  #3=\!dimenG
  \ignorespaces}

\def\!!Pythag{
  \!divide\!dimenE\!dimenH\!dimenI
  \advance\!dimenH by \!dimenI
    \divide\!dimenH by 2}

\def\placehypotenuse for <#1> and <#2> in <#3> {%
  \!Pythag{#1}{#2}{#3}}




\def\!qjoin (#1,#2) (#3,#4){%
  \advance\!intervalno by 1
  \!ifcoordmode
    \edef\!xmidpt{#1}\edef\!ymidpt{#2}%
  \else
    \!dimenA=#1\relax \edef\!xmidpt{\the\!dimenA}%
    \!dimenA=#2\relax \edef\!ymidpt{\the\!dimenA}%
  \fi
  \!xM=\!M{#1}\!xunit  \!yM=\!M{#2}\!yunit   \!rotateaboutpivot\!xM\!yM
  \!xE=\!M{#3}\!xunit  \!yE=\!M{#4}\!yunit   \!rotateaboutpivot\!xE\!yE
%
  \!dimenA=\!xM  \advance \!dimenA by -\!xS
  \!dimenB=\!xE  \advance \!dimenB by -\!xM
  \!xB=3\!dimenA \advance \!xB by -\!dimenB
  \!xC=2\!dimenB \advance \!xC by -2\!dimenA
%
  \!dimenA=\!yM  \advance \!dimenA by -\!yS%
  \!dimenB=\!yE  \advance \!dimenB by -\!yM%
  \!yB=3\!dimenA \advance \!yB by -\!dimenB%
  \!yC=2\!dimenB \advance \!yC by -2\!dimenA%
%
  \!xprime=\!xB  \!yprime=\!yB
  \!dxprime=.5\!xC  \!dyprime=.5\!yC
  \!getf \!midarclength=\!dimenA
  \!getf \advance \!midarclength by 4\!dimenA
  \!getf \advance \!midarclength by \!dimenA
  \divide \!midarclength by 12
%
  \!arclength=\!dimenA
  \!getf \advance \!arclength by 4\!dimenA
  \!getf \advance \!arclength by \!dimenA
  \divide \!arclength by 12
  \advance \!arclength by \!midarclength
  \global\advance \totalarclength by \!arclength
%
%
  \ifdim\!distacross>\!arclength 
    \advance \!distacross by -\!arclength
  \else
    \!initinverseinterp
    \loop\ifdim\!distacross<\!arclength
      \!inverseinterp
      \!xpos=\!t\!xC \advance\!xpos by \!xB
        \!xpos=\!t\!xpos \advance \!xpos by \!xS
      \!ypos=\!t\!yC \advance\!ypos by \!yB
        \!ypos=\!t\!ypos \advance \!ypos by \!yS
      \!plotifinbounds
      \advance\!distacross \plotsymbolspacing
      \!advancedashing
    \repeat  
    \advance \!distacross by -\!arclength
  \fi
  \!xS=\!xE
  \!yS=\!yE
  \ignorespaces}

\def\!getf{\!Pythag\!xprime\!yprime\!dimenA%
  \advance\!xprime by \!dxprime
  \advance\!yprime by \!dyprime}

\def\!initinverseinterp{%
  \ifdim\!arclength>\!zpt
    \!divide{8\!midarclength}\!arclength\!dimenE
    \ifdim\!dimenE<\!wmin \!setinverselinear
    \else 
      \ifdim\!dimenE>\!wmax \!setinverselinear
      \else
        \def\!inverseinterp{\!inversequad}\ignorespaces
%
%
         \!removept\!dimenE\!Ew
         \!dimenF=-\!Ew\!dimenE
         \advance\!dimenF by 32pt
         \!dimenG=8pt 
         \advance\!dimenG by -\!dimenE
         \!dimenG=\!Ew\!dimenG
         \!divide\!dimenF\!dimenG\!beta
         \!gamma=1pt
         \advance \!gamma by -\!beta
      \fi
    \fi
  \fi
  \ignorespaces}

\def\!inversequad{%
  \!divide\!distacross\!arclength\!dimenG
  \!removept\!dimenG\!v
  \!dimenG=\!v\!gamma
  \advance\!dimenG by \!beta
  \!dimenG=\!v\!dimenG
  \!removept\!dimenG\!t}

\def\!setinverselinear{%
  \def\!inverseinterp{\!inverselinear}%
  \divide\!dimenE by 8 \!removept\!dimenE\!t
  \!countC=\!intervalno \multiply \!countC 2
  \!countB=\!countC     \advance \!countB -1
  \!countA=\!countB     \advance \!countA -1
  \wlog{\the\!countB th point (\!xmidpt,\!ymidpt) being plotted 
    doesn't lie in the}%
  \wlog{ middle third of the arc between the \the\!countA th 
    and \the\!countC th points:}%
  \wlog{ [arc length \the\!countA\space to \the\!countB]/[arc length 
    \the \!countA\space to \the\!countC]=\!t.}%
  \ignorespaces}

\def\!inverselinear{%
  \!divide\!distacross\!arclength\!dimenG
  \!removept\!dimenG\!t}



\def\startrotation{%
  \let\!rotateaboutpivot=\!!rotateaboutpivot
  \let\!rotateonly=\!!rotateonly
  \!ifnextchar{b}{\!getsincos }%
    {\!getsincos by {\!cosrotationangle} {\!sinrotationangle} }}
\def\!getsincos by #1 #2 {%
  \edef\!cosrotationangle{#1}%
  \edef\!sinrotationangle{#2}%
  \!ifcoordmode 
    \let\!ROnext=\!ccheckforpivot
  \else
    \let\!ROnext=\!dcheckforpivot
  \fi
  \!ROnext}
\def\!ccheckforpivot{%
  \!ifnextchar{a}{\!cgetpivot}%
    {\!cgetpivot about {\!xpivotcoord} {\!ypivotcoord} }}
\def\!cgetpivot about #1 #2 {%
  \edef\!xpivotcoord{#1}%
  \edef\!ypivotcoord{#2}%
  \!xpivot=#1\!xunit  \!ypivot=#2\!yunit
  \ignorespaces}
\def\!dcheckforpivot{%
  \!ifnextchar{a}{\!dgetpivot}{\ignorespaces}}
\def\!dgetpivot about #1 #2 {%
  \!xpivot=#1\relax  \!ypivot=#2\relax
  \ignorespaces}

\def\stoprotation{%
  \let\!rotateaboutpivot=\!!!rotateaboutpivot
  \let\!rotateonly=\!!!rotateonly
  \ignorespaces}

\def\!!rotateaboutpivot#1#2{%
  \!dimenA=#1\relax  \advance\!dimenA -\!xpivot
  \!dimenB=#2\relax  \advance\!dimenB -\!ypivot
  \!dimenC=\!cosrotationangle\!dimenA
    \advance \!dimenC -\!sinrotationangle\!dimenB
  \!dimenD=\!cosrotationangle\!dimenB
    \advance \!dimenD  \!sinrotationangle\!dimenA
  \advance\!dimenC \!xpivot  \advance\!dimenD \!ypivot
  #1=\!dimenC  #2=\!dimenD
  \ignorespaces}

\def\!!rotateonly#1#2{%
  \!dimenA=#1\relax  \!dimenB=#2\relax 
  \!dimenC=\!cosrotationangle\!dimenA
    \advance \!dimenC -\!rotsign\!sinrotationangle\!dimenB
  \!dimenD=\!cosrotationangle\!dimenB
    \advance \!dimenD  \!rotsign\!sinrotationangle\!dimenA
  #1=\!dimenC  #2=\!dimenD
  \ignorespaces}
\def\!rotsign{}
\def\!!!rotateaboutpivot#1#2{\relax}
\def\!!!rotateonly#1#2{\relax}
\stoprotation

\def\!reverserotateonly#1#2{%
  \def\!rotsign{-}%
  \!rotateonly{#1}{#2}%
  \def\!rotsign{}%
  \ignorespaces}

\def\setshadegrid{%
  \!ifnextchar{s}{\!getspan }
    {\!getspan span <\!dshade>}}
\def\!getspan span <#1>{%
  \!dshade=#1\relax
  \!ifcoordmode 
    \let\!GRnext=\!GRccheckforAP
  \else
    \let\!GRnext=\!GRdcheckforAP
  \fi
  \!GRnext}
\def\!GRccheckforAP{%
  \!ifnextchar{p}{\!cgetanchor }
    {\!cgetanchor point at {\!xshadesave} {\!yshadesave} }}
\def\!cgetanchor point at #1 #2 {%
  \edef\!xshadesave{#1}\edef\!yshadesave{#2}%
  \!xshade=\!xshadesave\!xunit  \!yshade=\!yshadesave\!yunit
  \ignorespaces}
\def\!GRdcheckforAP{%
  \!ifnextchar{p}{\!dgetanchor}%
    {\ignorespaces}}
\def\!dgetanchor point at #1 #2 {%
  \!xshade=#1\relax  \!yshade=#2\relax
  \ignorespaces}

\def\setshadesymbol{%
  \!ifnextchar<{\!setshadesymbol}{\!setshadesymbol<,,,> }}

\def\!setshadesymbol <#1,#2,#3,#4> (#5#6){%
  \!setputobject{#5}{#6}%
  \setbox\!shadesymbol=\box\!putobject%
  \!shadesymbolxshift=\!xshift \!shadesymbolyshift=\!yshift
%
  \!dimenA=\!xshift \advance\!dimenA \!smidge
  \!override\!dimenA{#1}\!lshrinkage%
  \!dimenA=\!wd \advance \!dimenA -\!xshift
    \advance\!dimenA \!smidge
    \!override\!dimenA{#2}\!rshrinkage
  \!dimenA=\!dp \advance \!dimenA \!yshift
    \advance\!dimenA \!smidge
    \!override\!dimenA{#3}\!bshrinkage
  \!dimenA=\!ht \advance \!dimenA -\!yshift
    \advance\!dimenA \!smidge
    \!override\!dimenA{#4}\!tshrinkage
  \ignorespaces}
\def\!smidge{-.2pt}%

\def\!override#1#2#3{%
  \edef\!!override{#2}%
  \ifx \!!override\empty
    #3=#1\relax
  \else
    \if z\!!override
      #3=\!zpt
    \else
      \ifx \!!override\!blankz
        #3=\!zpt
      \else
        #3=#2\relax
      \fi
    \fi
  \fi
  \ignorespaces}
\def\!blankz{ z}


\def\!startvshade#1(#2,#3,#4){%
  \let\!!xunit=\!xunit%
  \let\!!yunit=\!yunit%
  \let\!!xshade=\!xshade%
  \let\!!yshade=\!yshade%
  \def\!getshrinkages{\!vgetshrinkages}%
  \let\!setshadelocation=\!vsetshadelocation%
  \!xS=\!M{#2}\!!xunit
  \!ybS=\!M{#3}\!!yunit
  \!ytS=\!M{#4}\!!yunit
  \!shadexorigin=\!xorigin  \advance \!shadexorigin \!shadesymbolxshift
  \!shadeyorigin=\!yorigin  \advance \!shadeyorigin \!shadesymbolyshift
  \ignorespaces}

\def\!starthshade#1(#2,#3,#4){%
  \let\!!xunit=\!yunit%
  \let\!!yunit=\!xunit%
  \let\!!xshade=\!yshade%
  \let\!!yshade=\!xshade%
  \def\!getshrinkages{\!hgetshrinkages}%
  \let\!setshadelocation=\!hsetshadelocation%
  \!xS=\!M{#2}\!!xunit
  \!ybS=\!M{#3}\!!yunit
  \!ytS=\!M{#4}\!!yunit
  \!shadexorigin=\!xorigin  \advance \!shadexorigin \!shadesymbolxshift
  \!shadeyorigin=\!yorigin  \advance \!shadeyorigin \!shadesymbolyshift
  \ignorespaces}

\def\!lattice#1#2#3#4#5{%
  \!dimenA=#1
  \!dimenB=#2
  \!countB=\!dimenB
%
  \!dimenC=#3
  \advance\!dimenC -\!dimenA
  \!countA=\!dimenC
  \divide\!countA \!countB
  \ifdim\!dimenC>\!zpt
    \!dimenD=\!countA\!dimenB
    \ifdim\!dimenD<\!dimenC
      \advance\!countA 1 
    \fi
  \fi
  \!dimenC=\!countA\!dimenB
    \advance\!dimenC \!dimenA
  #4=\!countA
  #5=\!dimenC
  \ignorespaces}

\def\!qshade#1(#2,#3,#4)#5(#6,#7,#8){%
  \!xM=\!M{#2}\!!xunit
  \!ybM=\!M{#3}\!!yunit
  \!ytM=\!M{#4}\!!yunit
  \!xE=\!M{#6}\!!xunit
  \!ybE=\!M{#7}\!!yunit
  \!ytE=\!M{#8}\!!yunit
  \!getcoeffs\!xS\!ybS\!xM\!ybM\!xE\!ybE\!ybB\!ybC
  \!getcoeffs\!xS\!ytS\!xM\!ytM\!xE\!ytE\!ytB\!ytC
  \def\!getylimits{\!qgetylimits}%
  \!shade{#1}\ignorespaces}

\def\!lshade#1(#2,#3,#4){%
  \!xE=\!M{#2}\!!xunit
  \!ybE=\!M{#3}\!!yunit
  \!ytE=\!M{#4}\!!yunit
  \!dimenE=\!xE  \advance \!dimenE -\!xS
  \!dimenC=\!ytE \advance \!dimenC -\!ytS
  \!divide\!dimenC\!dimenE\!ytB
  \!dimenC=\!ybE \advance \!dimenC -\!ybS
  \!divide\!dimenC\!dimenE\!ybB
  \def\!getylimits{\!lgetylimits}%
  \!shade{#1}\ignorespaces}

\def\!getcoeffs#1#2#3#4#5#6#7#8{%
  \!dimenC=#4\advance \!dimenC -#2
  \!dimenE=#3\advance \!dimenE -#1
  \!divide\!dimenC\!dimenE\!dimenF
  \!dimenC=#6\advance \!dimenC -#4
  \!dimenH=#5\advance \!dimenH -#3
  \!divide\!dimenC\!dimenH\!dimenG
  \advance\!dimenG -\!dimenF
  \advance \!dimenH \!dimenE
  \!divide\!dimenG\!dimenH#8
  \!removept#8\!t
  #7=-\!t\!dimenE
  \advance #7\!dimenF
  \ignorespaces}

\def\!shade#1{%
  \!getshrinkages#1<,,,>\!nil
  \advance \!dimenE \!xS
  \!lattice\!!xshade\!dshade\!dimenE
    \!parity\!xpos
  \!dimenF=-\!dimenF
    \advance\!dimenF \!xE
  \!loop\!not{\ifdim\!xpos>\!dimenF}
    \!shadecolumn%
    \advance\!xpos \!dshade
    \advance\!parity 1
  \repeat
  \!xS=\!xE
  \!ybS=\!ybE
  \!ytS=\!ytE
  \ignorespaces}

\def\!vgetshrinkages#1<#2,#3,#4,#5>#6\!nil{%
  \!override\!lshrinkage{#2}\!dimenE
  \!override\!rshrinkage{#3}\!dimenF
  \!override\!bshrinkage{#4}\!dimenG
  \!override\!tshrinkage{#5}\!dimenH
  \ignorespaces}
\def\!hgetshrinkages#1<#2,#3,#4,#5>#6\!nil{%
  \!override\!lshrinkage{#2}\!dimenG
  \!override\!rshrinkage{#3}\!dimenH
  \!override\!bshrinkage{#4}\!dimenE
  \!override\!tshrinkage{#5}\!dimenF
  \ignorespaces}

\def\!shadecolumn{%
  \!dxpos=\!xpos
  \advance\!dxpos -\!xS
  \!removept\!dxpos\!dx
  \!getylimits
  \advance\!ytpos -\!dimenH
  \advance\!ybpos \!dimenG
  \!yloc=\!!yshade
  \ifodd\!parity 
     \advance\!yloc \!dshade
  \fi
  \!lattice\!yloc{2\!dshade}\!ybpos%
    \!countA\!ypos
  \!dimenA=-\!shadexorigin \advance \!dimenA \!xpos
  \loop\!not{\ifdim\!ypos>\!ytpos}
    \!setshadelocation
    \!rotateaboutpivot\!xloc\!yloc%
    \!dimenA=-\!shadexorigin \advance \!dimenA \!xloc
    \!dimenB=-\!shadeyorigin \advance \!dimenB \!yloc
    \kern\!dimenA \raise\!dimenB\copy\!shadesymbol \kern-\!dimenA
    \advance\!ypos 2\!dshade
  \repeat
  \ignorespaces}

\def\!qgetylimits{%
  \!dimenA=\!dx\!ytC              
  \advance\!dimenA \!ytB
  \!ytpos=\!dx\!dimenA
  \advance\!ytpos \!ytS
  \!dimenA=\!dx\!ybC              
  \advance\!dimenA \!ybB
  \!ybpos=\!dx\!dimenA
  \advance\!ybpos \!ybS}

\def\!lgetylimits{%
  \!ytpos=\!dx\!ytB
  \advance\!ytpos \!ytS
  \!ybpos=\!dx\!ybB
  \advance\!ybpos \!ybS}

\def\!vsetshadelocation{
  \!xloc=\!xpos
  \!yloc=\!ypos}
\def\!hsetshadelocation{
  \!xloc=\!ypos
  \!yloc=\!xpos}





\def\!axisticks {%
  \def\!nextkeyword##1 {%
    \expandafter\ifx\csname !ticks##1\endcsname \relax
      \def\!next{\!fixkeyword{##1}}%
    \else
      \def\!next{\csname !ticks##1\endcsname}%
    \fi
    \!next}%
  \!axissetup
    \def\!axissetup{\relax}%
  \edef\!ticksinoutsign{\!ticksinoutSign}%
  \!ticklength=\longticklength
  \!tickwidth=\linethickness
  \!gridlinestatus
  \!setticktransform
  \!maketick
  \!tickcase=0
  \def\!LTlist{}%
  \!nextkeyword}

\def\ticksout{%
  \def\!ticksinoutSign{+}}
\def\ticksin{%
  \def\!ticksinoutSign{-}}
\ticksout

\def\gridlines{%
  \def\!gridlinestatus{\!gridlinestootrue}}
\def\nogridlines{%
  \def\!gridlinestatus{\!gridlinestoofalse}}
\nogridlines

\def\loggedticks{%
  \def\!setticktransform{\let\!ticktransform=\!logten}}
\def\unloggedticks{%
  \def\!setticktransform{\let\!ticktransform=\!donothing}}
\def\!donothing#1#2{\def#2{#1}}
\unloggedticks

\expandafter\def\csname !ticks/\endcsname{%
  \!not {\ifx \!LTlist\empty}
    \!placetickvalues
  \fi
  \def\!tickvalueslist{}%
  \def\!LTlist{}%
  \expandafter\csname !axis/\endcsname}

\def\!maketick{%
  \setbox\!boxA=\hbox{%
    \beginpicture
      \!setdimenmode
      \setcoordinatesystem point at {\!zpt} {\!zpt}   
      \linethickness=\!tickwidth
      \ifdim\!ticklength>\!zpt
        \putrule from {\!zpt} {\!zpt} to
          {\!ticksinoutsign\!tickxsign\!ticklength}
          {\!ticksinoutsign\!tickysign\!ticklength}
      \fi
      \if!gridlinestoo
        \putrule from {\!zpt} {\!zpt} to
          {-\!tickxsign\!xaxislength} {-\!tickysign\!yaxislength}
      \fi
    \endpicturesave <\!Xsave,\!Ysave>}%
    \wd\!boxA=\!zpt}
  
\def\!ticksin{%
  \def\!ticksinoutsign{-}%
  \!maketick
  \!nextkeyword}

\def\!ticksout{%
  \def\!ticksinoutsign{+}%
  \!maketick
  \!nextkeyword}

\def\!tickslength<#1> {%
  \!ticklength=#1\relax
  \!maketick
  \!nextkeyword}

\def\!tickslong{%
  \!tickslength<\longticklength> }

\def\!ticksshort{%
  \!tickslength<\shortticklength> }

\def\!tickswidth<#1> {%
  \!tickwidth=#1\relax
  \!maketick
  \!nextkeyword}

\def\!ticksandacross{%
  \!gridlinestootrue
  \!maketick
  \!nextkeyword}

\def\!ticksbutnotacross{%
  \!gridlinestoofalse
  \!maketick
  \!nextkeyword}

\def\!tickslogged{%
  \let\!ticktransform=\!logten
  \!nextkeyword}

\def\!ticksunlogged{%
  \let\!ticktransform=\!donothing
  \!nextkeyword}

\def\!ticksunlabeled{%
  \!tickcase=0
  \!nextkeyword}

\def\!ticksnumbered{%
  \!tickcase=1
  \!nextkeyword}

\def\!tickswithvalues#1/ {%
  \edef\!tickvalueslist{#1! /}%
  \!tickcase=2
  \!nextkeyword}

\def\!ticksquantity#1 {%
  \ifnum #1>1
    \!updatetickoffset
    \!countA=#1\relax
    \advance \!countA -1
    \!ticklocationincr=\!axisLength
      \divide \!ticklocationincr \!countA
    \!ticklocation=\!axisstart
    \loop \!not{\ifdim \!ticklocation>\!axisend}
      \!placetick\!ticklocation
      \ifcase\!tickcase
          \relax 
        \or
          \relax 
        \or
          \expandafter\!gettickvaluefrom\!tickvalueslist
          \edef\!tickfield{{\the\!ticklocation}{\!value}}%
          \expandafter\!listaddon\expandafter{\!tickfield}\!LTlist%
      \fi
      \advance \!ticklocation \!ticklocationincr
    \repeat
  \fi
  \!nextkeyword}

\def\!ticksat#1 {%
  \!updatetickoffset
  \edef\!Loc{#1}%
  \if /\!Loc
    \def\next{\!nextkeyword}%
  \else
    \!ticksincommon
    \def\next{\!ticksat}%
  \fi
  \next}    
      
\def\!ticksfrom#1 to #2 by #3 {%
  \!updatetickoffset
  \edef\!arg{#3}%
  \expandafter\!separate\!arg\!nil
  \!scalefactor=1
  \expandafter\!countfigures\!arg/
  \edef\!arg{#1}%
  \!scaleup\!arg by\!scalefactor to\!countE
  \edef\!arg{#2}%
  \!scaleup\!arg by\!scalefactor to\!countF
  \edef\!arg{#3}%
  \!scaleup\!arg by\!scalefactor to\!countG
  \loop \!not{\ifnum\!countE>\!countF}
    \ifnum\!scalefactor=1
      \edef\!Loc{\the\!countE}%
    \else
      \!scaledown\!countE by\!scalefactor to\!Loc
    \fi
    \!ticksincommon
    \advance \!countE \!countG
  \repeat
  \!nextkeyword}

\def\!updatetickoffset{%
  \!dimenA=\!ticksinoutsign\!ticklength
  \ifdim \!dimenA>\!offset
    \!offset=\!dimenA
  \fi}

\def\!placetick#1{%
  \if!xswitch
    \!xpos=#1\relax
    \!ypos=\!axisylevel
  \else
    \!xpos=\!axisxlevel
    \!ypos=#1\relax
  \fi
  \advance\!xpos \!Xsave
  \advance\!ypos \!Ysave
  \kern\!xpos\raise\!ypos\copy\!boxA\kern-\!xpos
  \ignorespaces}

\def\!gettickvaluefrom#1 #2 /{%
  \edef\!value{#1}%
  \edef\!tickvalueslist{#2 /}%
  \ifx \!tickvalueslist\!endtickvaluelist
    \!tickcase=0
  \fi}
\def\!endtickvaluelist{! /}

\def\!ticksincommon{%
  \!ticktransform\!Loc\!t
  \!ticklocation=\!t\!!unit
  \advance\!ticklocation -\!!origin
  \!placetick\!ticklocation
  \ifcase\!tickcase
    \relax 
  \or 
    \ifdim\!ticklocation<-\!!origin
      \edef\!Loc{$\!Loc$}%
    \fi
    \edef\!tickfield{{\the\!ticklocation}{\!Loc}}%
    \expandafter\!listaddon\expandafter{\!tickfield}\!LTlist%
  \or 
    \expandafter\!gettickvaluefrom\!tickvalueslist
    \edef\!tickfield{{\the\!ticklocation}{\!value}}%
    \expandafter\!listaddon\expandafter{\!tickfield}\!LTlist%
  \fi}

\def\!separate#1\!nil{%
  \!ifnextchar{-}{\!!separate}{\!!!separate}#1\!nil}
\def\!!separate-#1\!nil{%
  \def\!sign{-}%
  \!!!!separate#1..\!nil}
\def\!!!separate#1\!nil{%
  \def\!sign{+}%
  \!!!!separate#1..\!nil}
\def\!!!!separate#1.#2.#3\!nil{%
  \def\!arg{#1}%
  \ifx\!arg\!empty
    \!countA=0
  \else
    \!countA=\!arg
  \fi
  \def\!arg{#2}%
  \ifx\!arg\!empty
    \!countB=0
  \else
    \!countB=\!arg
  \fi}

\def\!countfigures#1{%
  \if #1/%
    \def\!next{\ignorespaces}%
  \else
    \multiply\!scalefactor 10
    \def\!next{\!countfigures}%
  \fi
  \!next}

\def\!scaleup#1by#2to#3{%
  \expandafter\!separate#1\!nil
  \multiply\!countA #2\relax
  \advance\!countA \!countB
  \if -\!sign
    \!countA=-\!countA
  \fi
  #3=\!countA
  \ignorespaces}

\def\!scaledown#1by#2to#3{%
  \!countA=#1\relax
  \ifnum \!countA<0 
    \def\!sign{-}
    \!countA=-\!countA
  \else
    \def\!sign{}%
  \fi
  \!countB=\!countA
  \divide\!countB #2\relax
  \!countC=\!countB
    \multiply\!countC #2\relax
  \advance \!countA -\!countC
  \edef#3{\!sign\the\!countB.}
  \!countC=\!countA 
  \ifnum\!countC=0 
    \!countC=1
  \fi
  \multiply\!countC 10
  \!loop \ifnum #2>\!countC
    \edef#3{#3\!zero}%
    \multiply\!countC 10
  \repeat
  \edef#3{#3\the\!countA}
  \ignorespaces}

\def\!placetickvalues{%
  \advance\!offset \tickstovaluesleading
  \if!xswitch
    \setbox\!boxA=\hbox{%
      \def\\##1##2{%
        \!dimenput {##2} [B] (##1,\!axisylevel)}%
      \beginpicture 
        \!LTlist
      \endpicturesave <\!Xsave,\!Ysave>}%
    \!dimenA=\!axisylevel
      \advance\!dimenA -\!Ysave
      \advance\!dimenA \!tickysign\!offset
      \if -\!tickysign
        \advance\!dimenA -\ht\!boxA
      \else
        \advance\!dimenA  \dp\!boxA
      \fi
    \advance\!offset \ht\!boxA 
      \advance\!offset \dp\!boxA
    \!dimenput {\box\!boxA} [Bl] <\!Xsave,\!Ysave> (\!zpt,\!dimenA)
  \else
    \setbox\!boxA=\hbox{%
      \def\\##1##2{%
        \!dimenput {##2} [r] (\!axisxlevel,##1)}%
      \beginpicture 
        \!LTlist
      \endpicturesave <\!Xsave,\!Ysave>}%
    \!dimenA=\!axisxlevel
      \advance\!dimenA -\!Xsave
      \advance\!dimenA \!tickxsign\!offset
      \if -\!tickxsign
        \advance\!dimenA -\wd\!boxA
      \fi
    \advance\!offset \wd\!boxA
    \!dimenput {\box\!boxA} [Bl] <\!Xsave,\!Ysave> (\!dimenA,\!zpt)
  \fi}

\normalgraphs
\catcode`!=12 


\tighten

\maketitle
\title{Some brane theoretic no-hair results (and their field theory duals)}
\author{Donald Marolf}
\address{Institute for Theoretical Physics, University of California,
Santa Barbara, CA 93106 \newline
Physics Department, Syracuse University, Syracuse,
         NY 13244}

\maketitle

\begin{abstract}
This contribution to the proceedings of the 1999 Canadian Conference
on General Relativity and Relativistic Astrophysics is a brief
exposition of earlier work, with Sumati Surya (hep-th/9805121) Amanda Peet
(hep-th/9903213), addressing certain results in higher dimensional
supergravity that are related to black hole no-hair theorems.
Its purpose is to describe, in language appropriate for
an audience of relativists, how these results can be related to the
Maldacena conjecture (aka, the AdS/CFT correspondence).  The end product
may be taken as a new kind of quantitative
evidence in support of the Maldacena conjecture.
\end{abstract}

\section{Introduction}
\label{intro}

The Maldacena conjecture \cite{Juan} (aka the AdS/CFT correspondence) has
been a topic of much interest and discussion in most communities
with interests in quantum gravity.  Here we provide some
commentary on a few recent results which, in the end, 
provide a new type of quantitative check on this
conjecture and its relatives \cite{IMSY}, 
beyond those previously known (see \cite{rev} for a review).  These
new results relate to the effective `delocalization' of charge
near a black hole horizon, a phenomenon associated with 
black hole no-hair theorems.  After discussing this phenomenon
in the familiar context of 3+1 Einstein-Maxwell theory,
we describe a related feature of ten-dimensional
supergravity, which can then be related to the Maldacena conjecture.
We will find that a corresponding phenomenon occurs in
the so-called dual field theory, and that the supergravity and
field theory results match at both the qualitative and quantitative
levels.

Due to a shortage of space, both citations of 
the literature and inclusion of technical details will be 
minimal.  In particular, as the intention is to make this paper
accessible to newcomers to string theory
and the Maldacena conjecture, stringy (and super Yang-Mills)
details will be particularly sparse, and true string/field theorists
are encouraged to go directly to the original works \cite{SM,MaPe}.
The goal of this paper is merely to provide a rough feel for the
results and, perhaps, to motivate the reader to examine the
original works.

\section{On No-Hair results}
\label{nohair}

Let us begin with a brief reminder of certain results associated with black hole
no-hair theorems.  For definiteness, consider Einstein-Maxwell theory (in 3+1 dimensions)
in the presence of charged dust.  Then we know that all stationary black hole solutions are
parameterized by their mass $M$, their charge $Q$, and their angular momentum $J$.
In particular, if $J=0$ then the solution is spherically symmetric.  

Now suppose that we take a bit of charged dust and drop it into a black hole.
The result will be a new black hole which will eventually settle down to a stationary state.
In particular, if both the black hole and the bit of dust had $J=0$, the result would be
spherically symmetric.  Thus, even though the bit of dust approaches 
from one side of the black hole,
the electric field becomes spherically symmetric in the
far future.  As the charge approaches the black hole horizon, the electric field
begins to become spherically symmetric due to the fact that the spacetime curvature
will bend the electric field lines around the black hole.  The result is that
the electric charge appears (when viewed from far away) to be `spread out' over
the horizon of the black hole.

In fact, this effect is not sensitive
to whether the electric charge actually falls
through the horizon.   Let us suppose that, at some point, the
charge is attached
to a powerful rocket that keeps it from falling further into the black hole, 
perhaps moving
instead along one of the worldlines shown in the conformal diagram
on the left below.   The
curvature of spacetime 
bends the electric field lines and tends to make the
electric field spherically symmetric.  A quasi-artistic
impression of this process is shown in the diagram on the right
below.  There, the small dot denotes the bit of charge and the lines
describe the electric field produced by the charge.
The extent to which the electric field
appears spherical is determined by just how close to the horizon  the charge
actually sits.  In the limit in which the charge approaches
the horizon, the electric field becomes spherically symmetric.

As a final conceptual jump, let us dispense with the rockets used above 
by passing to the extremal case, in which both the charged dust and the black
hole satisfy $Q=+M$.
Now, the electrostatic repulsion alone is sufficient to keep the charge from
falling into the black hole.  We could consider a family, labeled
by a parameter $\Delta$ as shown on the left below, of such
extremal charges which do not fall toward the black hole at all.  Instead, they
sit close to an extremal hole following
the integral curves of a timelike killing vector field.  Once again, in the limit where
the worldline of the charge approaches the horizon, the electric field becomes
spherically symmetric and the charge appears to have spread out over the horizon
of the black hole.  We will refer to this effect as ``delocalization'' of the 
charge.  Note that the diagram on the right
below shows only the field lines produced by the additional charge, and
not the field lines produced by the black hole itself.


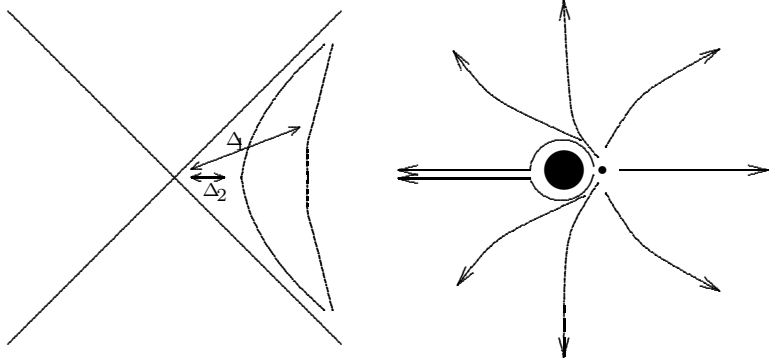
\font\thinlinefont=cmr5
\begingroup\makeatletter\ifx\SetFigFont\undefined
\def\x#1#2#3#4#5#6#7\relax{\def\x{#1#2#3#4#5#6}}%
\expandafter\x\fmtname xxxxxx\relax \def\y{splain}%
\ifx\x\y   
\gdef\SetFigFont#1#2#3{%
  \ifnum #1<17\tiny\else \ifnum #1<20\small\else
  \ifnum #1<24\normalsize\else \ifnum #1<29\large\else
  \ifnum #1<34\Large\else \ifnum #1<41\LARGE\else
     \huge\fi\fi\fi\fi\fi\fi
  \csname #3\endcsname}%
\else
\gdef\SetFigFont#1#2#3{\begingroup
  \count@#1\relax \ifnum 25<\count@\count@25\fi
  \def\x{\endgroup\@setsize\SetFigFont{#2pt}}%
  \expandafter\x
    \csname \romannumeral\the\count@ pt\expandafter\endcsname
    \csname @\romannumeral\the\count@ pt\endcsname
  \csname #3\endcsname}%
\fi
\fi\endgroup
\mbox{\beginpicture
\setcoordinatesystem units <1.00000cm,1.00000cm>
\unitlength=1.00000cm
\linethickness=1pt
\setplotsymbol ({\makebox(0,0)[l]{\tencirc\symbol{'160}}})
\setshadesymbol ({\thinlinefont .})
\setlinear
%
%
\linethickness= 0.500pt
\setplotsymbol ({\thinlinefont .})
\circulararc 166.378 degrees from 15.725 21.395 center at 15.305 21.316
%
%
\linethickness= 0.500pt
\setplotsymbol ({\thinlinefont .})
\circulararc 148.556 degrees from 14.878 21.226 center at 15.293 21.374
%
%
\linethickness= 0.500pt
\setplotsymbol ({\thinlinefont .})
\put{\makebox(0,0)[l]{\circle*{ 0.114}}} at 15.837 21.338
%
%
\linethickness= 0.500pt
\setplotsymbol ({\thinlinefont .})
\put{\makebox(0,0)[l]{\circle*{ 0.686}}} at 15.327 21.338
%
%
\linethickness= 0.500pt
\setplotsymbol ({\thinlinefont .})
\putrule from 16.059 21.338 to 18.040 21.338
%
%
\plot 17.786 21.275 18.040 21.338 17.786 21.402 /
%
%
%
\linethickness= 0.500pt
\setplotsymbol ({\thinlinefont .})
\putrule from 14.878 21.338 to 13.123 21.338
%
%
\plot 13.377 21.402 13.123 21.338 13.377 21.275 /
%
%
%
\linethickness= 0.500pt
\setplotsymbol ({\thinlinefont .})
\putrule from 14.878 21.226 to 13.123 21.226
%
%
\plot 13.377 21.289 13.123 21.226 13.377 21.162 /
%
%
%
\linethickness= 0.500pt
\setplotsymbol ({\thinlinefont .})
\plot  7.938 23.453 12.363 19.027 /
%
%
\linethickness= 0.500pt
\setplotsymbol ({\thinlinefont .})
\plot  7.938 19.027 12.363 23.453 /
%
%
\linethickness= 0.500pt
\setplotsymbol ({\thinlinefont .})
%
%
\plot 10.471 21.446 10.372 21.351 10.509 21.347 /
\plot 10.372 21.351 11.809 21.903 /
%
%
\plot 11.709 21.808 11.809 21.903 11.671 21.907 /
%
%
%
\linethickness= 0.500pt
\setplotsymbol ({\thinlinefont .})
%
%
\plot 10.499 21.294 10.372 21.241 10.499 21.188 /
\putrule from 10.372 21.241 to 10.814 21.241
%
%
\plot 10.687 21.188 10.814 21.241 10.687 21.294 /
%
%
%
\linethickness= 0.500pt
\setplotsymbol ({\thinlinefont .})
\plot 15.782 21.167 15.780 21.162 /
\plot 15.780 21.162 15.773 21.156 /
\plot 15.773 21.156 15.765 21.141 /
\plot 15.765 21.141 15.750 21.120 /
\plot 15.750 21.120 15.731 21.093 /
\plot 15.731 21.093 15.708 21.057 /
\plot 15.708 21.057 15.680 21.014 /
\plot 15.680 21.014 15.651 20.966 /
\plot 15.651 20.966 15.619 20.915 /
\plot 15.619 20.915 15.587 20.862 /
\plot 15.587 20.862 15.555 20.807 /
\plot 15.555 20.807 15.524 20.752 /
\plot 15.524 20.752 15.496 20.699 /
\plot 15.496 20.699 15.471 20.646 /
\plot 15.471 20.646 15.447 20.593 /
\plot 15.447 20.593 15.428 20.542 /
\plot 15.428 20.542 15.411 20.489 /
\plot 15.411 20.489 15.397 20.436 /
\plot 15.397 20.436 15.384 20.379 /
\plot 15.384 20.379 15.375 20.333 /
\plot 15.375 20.333 15.369 20.284 /
\plot 15.369 20.284 15.363 20.231 /
\plot 15.363 20.231 15.356 20.174 /
\plot 15.356 20.174 15.352 20.113 /
\plot 15.352 20.113 15.348 20.045 /
\plot 15.348 20.045 15.344 19.971 /
\plot 15.344 19.971 15.339 19.892 /
\plot 15.339 19.892 15.337 19.810 /
\plot 15.337 19.810 15.335 19.723 /
\plot 15.335 19.723 15.333 19.634 /
\plot 15.333 19.634 15.331 19.541 /
\putrule from 15.331 19.541 to 15.331 19.450
\plot 15.331 19.450 15.329 19.359 /
\putrule from 15.329 19.359 to 15.329 19.272
\putrule from 15.329 19.272 to 15.329 19.190
\plot 15.329 19.190 15.327 19.113 /
\putrule from 15.327 19.113 to 15.327 19.048
\putrule from 15.327 19.048 to 15.327 18.991
\putrule from 15.327 18.991 to 15.327 18.944
\putrule from 15.327 18.944 to 15.327 18.908
\putrule from 15.327 18.908 to 15.327 18.853
%
%
\plot 15.263 19.107 15.327 18.853 15.390 19.107 /
%
%
%
\linethickness= 0.500pt
\setplotsymbol ({\thinlinefont .})
\plot 15.782 21.507 15.778 21.512 /
\plot 15.778 21.512 15.769 21.520 /
\plot 15.769 21.520 15.756 21.535 /
\plot 15.756 21.535 15.735 21.558 /
\plot 15.735 21.558 15.710 21.588 /
\plot 15.710 21.588 15.678 21.626 /
\plot 15.678 21.626 15.646 21.666 /
\plot 15.646 21.666 15.613 21.711 /
\plot 15.613 21.711 15.579 21.755 /
\plot 15.579 21.755 15.547 21.802 /
\plot 15.547 21.802 15.519 21.850 /
\plot 15.519 21.850 15.496 21.901 /
\plot 15.496 21.901 15.475 21.952 /
\plot 15.475 21.952 15.456 22.009 /
\plot 15.456 22.009 15.441 22.070 /
\plot 15.441 22.070 15.433 22.113 /
\plot 15.433 22.113 15.424 22.157 /
\plot 15.424 22.157 15.418 22.208 /
\plot 15.418 22.208 15.411 22.263 /
\plot 15.411 22.263 15.403 22.324 /
\plot 15.403 22.324 15.397 22.390 /
\plot 15.397 22.390 15.392 22.462 /
\plot 15.392 22.462 15.386 22.540 /
\plot 15.386 22.540 15.380 22.625 /
\plot 15.380 22.625 15.373 22.712 /
\plot 15.373 22.712 15.369 22.803 /
\plot 15.369 22.803 15.363 22.896 /
\plot 15.363 22.896 15.356 22.989 /
\plot 15.356 22.989 15.352 23.080 /
\plot 15.352 23.080 15.348 23.169 /
\plot 15.348 23.169 15.344 23.254 /
\plot 15.344 23.254 15.339 23.330 /
\plot 15.339 23.330 15.335 23.398 /
\plot 15.335 23.398 15.333 23.455 /
\plot 15.333 23.455 15.331 23.503 /
\plot 15.331 23.503 15.329 23.539 /
\plot 15.329 23.539 15.327 23.597 /
%
%
\plot 15.400 23.345 15.327 23.597 15.273 23.340 /
%
%
%
\linethickness= 0.500pt
\setplotsymbol ({\thinlinefont .})
\plot 15.610 20.997 15.606 20.995 /
\plot 15.606 20.995 15.598 20.991 /
\plot 15.598 20.991 15.583 20.985 /
\plot 15.583 20.985 15.560 20.974 /
\plot 15.560 20.974 15.526 20.959 /
\plot 15.526 20.959 15.486 20.940 /
\plot 15.486 20.940 15.435 20.917 /
\plot 15.435 20.917 15.375 20.889 /
\plot 15.375 20.889 15.310 20.860 /
\plot 15.310 20.860 15.242 20.826 /
\plot 15.242 20.826 15.168 20.792 /
\plot 15.168 20.792 15.094 20.758 /
\plot 15.094 20.758 15.022 20.722 /
\plot 15.022 20.722 14.950 20.688 /
\plot 14.950 20.688 14.880 20.654 /
\plot 14.880 20.654 14.817 20.623 /
\plot 14.817 20.623 14.755 20.591 /
\plot 14.755 20.591 14.698 20.561 /
\plot 14.698 20.561 14.647 20.534 /
\plot 14.647 20.534 14.601 20.508 /
\plot 14.601 20.508 14.558 20.481 /
\plot 14.558 20.481 14.518 20.458 /
\plot 14.518 20.458 14.482 20.432 /
\plot 14.482 20.432 14.436 20.398 /
\plot 14.436 20.398 14.393 20.364 /
\plot 14.393 20.364 14.351 20.326 /
\plot 14.351 20.326 14.309 20.286 /
\plot 14.309 20.286 14.266 20.244 /
\plot 14.266 20.244 14.224 20.195 /
\plot 14.224 20.195 14.182 20.146 /
\plot 14.182 20.146 14.137 20.094 /
\plot 14.137 20.094 14.095 20.043 /
\plot 14.095 20.043 14.055 19.992 /
\plot 14.055 19.992 14.017 19.945 /
\plot 14.017 19.945 13.985 19.903 /
\plot 13.985 19.903 13.959 19.869 /
\plot 13.959 19.869 13.915 19.812 /
%
%
\plot 14.021 20.051 13.915 19.812 14.121 19.974 /
%
%
%
\linethickness= 0.500pt
\setplotsymbol ({\thinlinefont .})
\plot 15.555 21.736 15.551 21.738 /
\plot 15.551 21.738 15.543 21.742 /
\plot 15.543 21.742 15.528 21.749 /
\plot 15.528 21.749 15.505 21.759 /
\plot 15.505 21.759 15.471 21.774 /
\plot 15.471 21.774 15.428 21.793 /
\plot 15.428 21.793 15.378 21.816 /
\plot 15.378 21.816 15.320 21.844 /
\plot 15.320 21.844 15.255 21.874 /
\plot 15.255 21.874 15.185 21.905 /
\plot 15.185 21.905 15.111 21.939 /
\plot 15.111 21.939 15.037 21.973 /
\plot 15.037 21.973 14.963 22.007 /
\plot 14.963 22.007 14.891 22.043 /
\plot 14.891 22.043 14.823 22.075 /
\plot 14.823 22.075 14.757 22.109 /
\plot 14.757 22.109 14.696 22.138 /
\plot 14.696 22.138 14.641 22.168 /
\plot 14.641 22.168 14.588 22.195 /
\plot 14.588 22.195 14.542 22.223 /
\plot 14.542 22.223 14.499 22.248 /
\plot 14.499 22.248 14.459 22.272 /
\plot 14.459 22.272 14.423 22.297 /
\plot 14.423 22.297 14.376 22.331 /
\plot 14.376 22.331 14.334 22.365 /
\plot 14.334 22.365 14.292 22.403 /
\plot 14.292 22.403 14.252 22.443 /
\plot 14.252 22.443 14.209 22.487 /
\plot 14.209 22.487 14.167 22.534 /
\plot 14.167 22.534 14.122 22.585 /
\plot 14.122 22.585 14.080 22.636 /
\plot 14.080 22.636 14.038 22.689 /
\plot 14.038 22.689 13.998 22.739 /
\plot 13.998 22.739 13.962 22.786 /
\plot 13.962 22.786 13.930 22.828 /
\plot 13.930 22.828 13.902 22.862 /
\plot 13.902 22.862 13.860 22.919 /
%
%
\plot 14.062 22.753 13.860 22.919 13.960 22.677 /
%
%
%
\linethickness= 0.500pt
\setplotsymbol ({\thinlinefont .})
\plot 15.883 21.038 15.886 21.033 /
\plot 15.886 21.033 15.890 21.025 /
\plot 15.890 21.025 15.900 21.010 /
\plot 15.900 21.010 15.913 20.989 /
\plot 15.913 20.989 15.930 20.957 /
\plot 15.930 20.957 15.953 20.919 /
\plot 15.953 20.919 15.981 20.877 /
\plot 15.981 20.877 16.010 20.826 /
\plot 16.010 20.826 16.042 20.775 /
\plot 16.042 20.775 16.076 20.722 /
\plot 16.076 20.722 16.112 20.669 /
\plot 16.112 20.669 16.146 20.616 /
\plot 16.146 20.616 16.182 20.568 /
\plot 16.182 20.568 16.216 20.521 /
\plot 16.216 20.521 16.248 20.479 /
\plot 16.248 20.479 16.281 20.439 /
\plot 16.281 20.439 16.315 20.403 /
\plot 16.315 20.403 16.351 20.369 /
\plot 16.351 20.369 16.387 20.335 /
\plot 16.387 20.335 16.423 20.305 /
\plot 16.423 20.305 16.461 20.276 /
\plot 16.461 20.276 16.504 20.244 /
\plot 16.504 20.244 16.550 20.212 /
\plot 16.550 20.212 16.603 20.180 /
\plot 16.603 20.180 16.660 20.144 /
\plot 16.660 20.144 16.722 20.106 /
\plot 16.722 20.106 16.787 20.068 /
\plot 16.787 20.068 16.855 20.028 /
\plot 16.855 20.028 16.927 19.988 /
\plot 16.927 19.988 16.999 19.947 /
\plot 16.999 19.947 17.069 19.909 /
\plot 17.069 19.909 17.137 19.871 /
\plot 17.137 19.871 17.198 19.837 /
\plot 17.198 19.837 17.253 19.808 /
\plot 17.253 19.808 17.297 19.782 /
\plot 17.297 19.782 17.335 19.763 /
\plot 17.335 19.763 17.393 19.732 /
%
%
\plot 17.140 19.799 17.393 19.732 17.201 19.910 /
%
%
%
\linethickness= 0.500pt
\setplotsymbol ({\thinlinefont .})
\plot 15.883 21.643 15.886 21.647 /
\plot 15.886 21.647 15.890 21.656 /
\plot 15.890 21.656 15.900 21.670 /
\plot 15.900 21.670 15.913 21.692 /
\plot 15.913 21.692 15.930 21.723 /
\plot 15.930 21.723 15.953 21.761 /
\plot 15.953 21.761 15.981 21.804 /
\plot 15.981 21.804 16.010 21.855 /
\plot 16.010 21.855 16.042 21.905 /
\plot 16.042 21.905 16.076 21.958 /
\plot 16.076 21.958 16.112 22.011 /
\plot 16.112 22.011 16.146 22.064 /
\plot 16.146 22.064 16.182 22.113 /
\plot 16.182 22.113 16.216 22.159 /
\plot 16.216 22.159 16.248 22.202 /
\plot 16.248 22.202 16.281 22.242 /
\plot 16.281 22.242 16.315 22.278 /
\plot 16.315 22.278 16.351 22.312 /
\plot 16.351 22.312 16.387 22.346 /
\plot 16.387 22.346 16.423 22.375 /
\plot 16.423 22.375 16.461 22.405 /
\plot 16.461 22.405 16.504 22.437 /
\plot 16.504 22.437 16.550 22.468 /
\plot 16.550 22.468 16.603 22.500 /
\plot 16.603 22.500 16.660 22.536 /
\plot 16.660 22.536 16.722 22.574 /
\plot 16.722 22.574 16.787 22.612 /
\plot 16.787 22.612 16.855 22.653 /
\plot 16.855 22.653 16.927 22.693 /
\plot 16.927 22.693 16.999 22.733 /
\plot 16.999 22.733 17.069 22.771 /
\plot 17.069 22.771 17.137 22.809 /
\plot 17.137 22.809 17.198 22.843 /
\plot 17.198 22.843 17.253 22.873 /
\plot 17.253 22.873 17.297 22.898 /
\plot 17.297 22.898 17.335 22.917 /
\plot 17.335 22.917 17.393 22.949 /
%
%
\plot 17.201 22.770 17.393 22.949 17.140 22.881 /
%
%
%
\linethickness= 0.500pt
\setplotsymbol ({\thinlinefont .})
\putrule from 11.921 21.241 to 11.921 21.234
\putrule from 11.921 21.234 to 11.921 21.220
\plot 11.921 21.220 11.919 21.196 /
\putrule from 11.919 21.196 to 11.919 21.165
\plot 11.919 21.165 11.917 21.124 /
\plot 11.917 21.124 11.915 21.078 /
\putrule from 11.915 21.078 to 11.915 21.029
\plot 11.915 21.029 11.913 20.983 /
\putrule from 11.913 20.983 to 11.913 20.940
\plot 11.913 20.940 11.915 20.898 /
\putrule from 11.915 20.898 to 11.915 20.862
\plot 11.915 20.862 11.917 20.828 /
\plot 11.917 20.828 11.921 20.796 /
\plot 11.921 20.796 11.925 20.764 /
\plot 11.925 20.764 11.932 20.733 /
\plot 11.932 20.733 11.938 20.701 /
\plot 11.938 20.701 11.946 20.669 /
\plot 11.946 20.669 11.955 20.637 /
\plot 11.955 20.637 11.963 20.606 /
\plot 11.963 20.606 11.972 20.574 /
\plot 11.972 20.574 11.980 20.544 /
\plot 11.980 20.544 11.991 20.513 /
\plot 11.991 20.513 11.999 20.481 /
\plot 11.999 20.481 12.008 20.449 /
\plot 12.008 20.449 12.016 20.417 /
\plot 12.016 20.417 12.025 20.386 /
\plot 12.025 20.386 12.033 20.354 /
\plot 12.033 20.354 12.042 20.322 /
\plot 12.042 20.322 12.048 20.290 /
\plot 12.048 20.290 12.057 20.259 /
\plot 12.057 20.259 12.065 20.227 /
\plot 12.065 20.227 12.073 20.195 /
\plot 12.073 20.195 12.080 20.163 /
\plot 12.080 20.163 12.088 20.132 /
\plot 12.088 20.132 12.097 20.102 /
\plot 12.097 20.102 12.103 20.070 /
\plot 12.103 20.070 12.112 20.038 /
\plot 12.112 20.038 12.120 20.007 /
\plot 12.120 20.007 12.129 19.975 /
\plot 12.129 19.975 12.135 19.943 /
\plot 12.135 19.943 12.143 19.911 /
\plot 12.143 19.911 12.152 19.880 /
\plot 12.152 19.880 12.160 19.846 /
\plot 12.160 19.846 12.169 19.810 /
\plot 12.169 19.810 12.179 19.770 /
\plot 12.179 19.770 12.190 19.725 /
\plot 12.190 19.725 12.200 19.679 /
\plot 12.200 19.679 12.213 19.632 /
\plot 12.213 19.632 12.224 19.586 /
\plot 12.224 19.586 12.234 19.545 /
\plot 12.234 19.545 12.243 19.514 /
\plot 12.243 19.514 12.249 19.490 /
\plot 12.249 19.490 12.251 19.475 /
\plot 12.251 19.475 12.253 19.469 /
%
%
\linethickness= 0.500pt
\setplotsymbol ({\thinlinefont .})
\putrule from 11.921 21.241 to 11.921 21.247
\putrule from 11.921 21.247 to 11.921 21.262
\plot 11.921 21.262 11.919 21.285 /
\putrule from 11.919 21.285 to 11.919 21.317
\plot 11.919 21.317 11.917 21.357 /
\plot 11.917 21.357 11.915 21.404 /
\putrule from 11.915 21.404 to 11.915 21.450
\plot 11.915 21.450 11.913 21.497 /
\putrule from 11.913 21.497 to 11.913 21.541
\plot 11.913 21.541 11.915 21.582 /
\putrule from 11.915 21.582 to 11.915 21.618
\plot 11.915 21.618 11.917 21.651 /
\plot 11.917 21.651 11.921 21.683 /
\plot 11.921 21.683 11.925 21.715 /
\plot 11.925 21.715 11.932 21.747 /
\plot 11.932 21.747 11.938 21.778 /
\plot 11.938 21.778 11.946 21.810 /
\plot 11.946 21.810 11.955 21.842 /
\plot 11.955 21.842 11.963 21.874 /
\plot 11.963 21.874 11.972 21.905 /
\plot 11.972 21.905 11.980 21.935 /
\plot 11.980 21.935 11.991 21.967 /
\plot 11.991 21.967 11.999 21.999 /
\plot 11.999 21.999 12.008 22.030 /
\plot 12.008 22.030 12.016 22.062 /
\plot 12.016 22.062 12.025 22.094 /
\plot 12.025 22.094 12.033 22.126 /
\plot 12.033 22.126 12.042 22.157 /
\plot 12.042 22.157 12.048 22.189 /
\plot 12.048 22.189 12.057 22.221 /
\plot 12.057 22.221 12.065 22.253 /
\plot 12.065 22.253 12.073 22.284 /
\plot 12.073 22.284 12.080 22.316 /
\plot 12.080 22.316 12.088 22.348 /
\plot 12.088 22.348 12.097 22.377 /
\plot 12.097 22.377 12.103 22.409 /
\plot 12.103 22.409 12.112 22.441 /
\plot 12.112 22.441 12.120 22.473 /
\plot 12.120 22.473 12.129 22.504 /
\plot 12.129 22.504 12.135 22.536 /
\plot 12.135 22.536 12.143 22.568 /
\plot 12.143 22.568 12.152 22.600 /
\plot 12.152 22.600 12.160 22.634 /
\plot 12.160 22.634 12.169 22.669 /
\plot 12.169 22.669 12.179 22.710 /
\plot 12.179 22.710 12.190 22.754 /
\plot 12.190 22.754 12.200 22.801 /
\plot 12.200 22.801 12.213 22.847 /
\plot 12.213 22.847 12.224 22.894 /
\plot 12.224 22.894 12.234 22.934 /
\plot 12.234 22.934 12.243 22.966 /
\plot 12.243 22.966 12.249 22.989 /
\plot 12.249 22.989 12.251 23.004 /
\plot 12.251 23.004 12.253 23.010 /
%
%
\linethickness= 0.500pt
\setplotsymbol ({\thinlinefont .})
\plot 11.036 21.241 11.038 21.247 /
\plot 11.038 21.247 11.041 21.262 /
\plot 11.041 21.262 11.045 21.285 /
\plot 11.045 21.285 11.053 21.317 /
\plot 11.053 21.317 11.062 21.357 /
\plot 11.062 21.357 11.070 21.404 /
\plot 11.070 21.404 11.081 21.450 /
\plot 11.081 21.450 11.093 21.497 /
\plot 11.093 21.497 11.104 21.541 /
\plot 11.104 21.541 11.115 21.582 /
\plot 11.115 21.582 11.125 21.618 /
\plot 11.125 21.618 11.136 21.651 /
\plot 11.136 21.651 11.146 21.683 /
\plot 11.146 21.683 11.159 21.715 /
\plot 11.159 21.715 11.172 21.747 /
\plot 11.172 21.747 11.187 21.778 /
\plot 11.187 21.778 11.201 21.810 /
\plot 11.201 21.810 11.216 21.842 /
\plot 11.216 21.842 11.231 21.874 /
\plot 11.231 21.874 11.248 21.905 /
\plot 11.248 21.905 11.263 21.935 /
\plot 11.263 21.935 11.280 21.967 /
\plot 11.280 21.967 11.297 21.999 /
\plot 11.297 21.999 11.314 22.030 /
\plot 11.314 22.030 11.331 22.062 /
\plot 11.331 22.062 11.350 22.094 /
\plot 11.350 22.094 11.369 22.126 /
\plot 11.369 22.126 11.386 22.153 /
\plot 11.386 22.153 11.405 22.181 /
\plot 11.405 22.181 11.424 22.208 /
\plot 11.424 22.208 11.443 22.236 /
\plot 11.443 22.236 11.462 22.263 /
\plot 11.462 22.263 11.483 22.291 /
\plot 11.483 22.291 11.502 22.318 /
\plot 11.502 22.318 11.523 22.346 /
\plot 11.523 22.346 11.544 22.375 /
\plot 11.544 22.375 11.565 22.403 /
\plot 11.565 22.403 11.587 22.430 /
\plot 11.587 22.430 11.608 22.458 /
\plot 11.608 22.458 11.629 22.485 /
\plot 11.629 22.485 11.652 22.513 /
\plot 11.652 22.513 11.676 22.540 /
\plot 11.676 22.540 11.699 22.568 /
\plot 11.699 22.568 11.724 22.595 /
\plot 11.724 22.595 11.750 22.625 /
\plot 11.750 22.625 11.781 22.657 /
\plot 11.781 22.657 11.813 22.691 /
\plot 11.813 22.691 11.851 22.727 /
\plot 11.851 22.727 11.889 22.765 /
\plot 11.889 22.765 11.932 22.807 /
\plot 11.932 22.807 11.974 22.847 /
\plot 11.974 22.847 12.016 22.888 /
\plot 12.016 22.888 12.054 22.926 /
\plot 12.054 22.926 12.086 22.955 /
\plot 12.086 22.955 12.112 22.981 /
\plot 12.112 22.981 12.129 22.995 /
\plot 12.129 22.995 12.139 23.006 /
\plot 12.139 23.006 12.143 23.010 /
%
%
\linethickness= 0.500pt
\setplotsymbol ({\thinlinefont .})
\plot 11.036 21.241 11.038 21.234 /
\plot 11.038 21.234 11.041 21.220 /
\plot 11.041 21.220 11.045 21.196 /
\plot 11.045 21.196 11.053 21.165 /
\plot 11.053 21.165 11.062 21.124 /
\plot 11.062 21.124 11.070 21.078 /
\plot 11.070 21.078 11.081 21.029 /
\plot 11.081 21.029 11.093 20.983 /
\plot 11.093 20.983 11.104 20.940 /
\plot 11.104 20.940 11.115 20.898 /
\plot 11.115 20.898 11.125 20.862 /
\plot 11.125 20.862 11.136 20.828 /
\plot 11.136 20.828 11.146 20.796 /
\plot 11.146 20.796 11.159 20.764 /
\plot 11.159 20.764 11.172 20.733 /
\plot 11.172 20.733 11.187 20.701 /
\plot 11.187 20.701 11.201 20.669 /
\plot 11.201 20.669 11.216 20.637 /
\plot 11.216 20.637 11.231 20.606 /
\plot 11.231 20.606 11.248 20.574 /
\plot 11.248 20.574 11.263 20.544 /
\plot 11.263 20.544 11.280 20.513 /
\plot 11.280 20.513 11.297 20.481 /
\plot 11.297 20.481 11.314 20.449 /
\plot 11.314 20.449 11.331 20.417 /
\plot 11.331 20.417 11.350 20.386 /
\plot 11.350 20.386 11.369 20.354 /
\plot 11.369 20.354 11.386 20.326 /
\plot 11.386 20.326 11.405 20.299 /
\plot 11.405 20.299 11.424 20.271 /
\plot 11.424 20.271 11.443 20.244 /
\plot 11.443 20.244 11.462 20.216 /
\plot 11.462 20.216 11.483 20.189 /
\plot 11.483 20.189 11.502 20.161 /
\plot 11.502 20.161 11.523 20.132 /
\plot 11.523 20.132 11.544 20.104 /
\plot 11.544 20.104 11.565 20.077 /
\plot 11.565 20.077 11.587 20.049 /
\plot 11.587 20.049 11.608 20.022 /
\plot 11.608 20.022 11.629 19.994 /
\plot 11.629 19.994 11.652 19.967 /
\plot 11.652 19.967 11.676 19.939 /
\plot 11.676 19.939 11.699 19.911 /
\plot 11.699 19.911 11.724 19.884 /
\plot 11.724 19.884 11.750 19.854 /
\plot 11.750 19.854 11.781 19.823 /
\plot 11.781 19.823 11.813 19.789 /
\plot 11.813 19.789 11.851 19.753 /
\plot 11.851 19.753 11.889 19.715 /
\plot 11.889 19.715 11.932 19.672 /
\plot 11.932 19.672 11.974 19.632 /
\plot 11.974 19.632 12.016 19.592 /
\plot 12.016 19.592 12.054 19.554 /
\plot 12.054 19.554 12.086 19.524 /
\plot 12.086 19.524 12.112 19.499 /
\plot 12.112 19.499 12.129 19.484 /
\plot 12.129 19.484 12.139 19.473 /
\plot 12.139 19.473 12.143 19.469 /
%
%
\put{\SetFigFont{7}{8.4}{rm}$\Delta$} [lB] at 10.520 20.995
%
%
\put{\SetFigFont{7}{8.4}{rm}$\Delta$} [lB] at 10.827 21.670
%
%
\put{\SetFigFont{7}{8.4}{rm}2} [lB] at 10.704 20.932
%
%
\put{\SetFigFont{7}{8.4}{rm}1} [lB] at 10.950 21.609
\linethickness=0pt
\putrectangle corners at  7.912 23.622 and 18.066 18.828
\endpicture}


We will shortly be interested in analogous phenomena in 10-d
supergravity.  What we will do is to use some qualitative features of this
effect, along with certain aspects of the supergravity/field theory correspondence, to suggest
an analogue in the field theory.  Assuming
this guess to be correct, the field theory makes both qualitative predictions
about when effects of this sort should occur and also {\it quantitative}
predictions\footnote{The term `prediction' should perhaps be taken
with a grain of salt.  Even assuming both that the Maldacena conjecture is
correct and that the field theory analogue of our effect
has been correctly guessed, it is not a priori obvious that naive calculations
of the type that we will discuss must be quantitatively correct.
Nonetheless, that such naive results will in fact precisely
agree with the supergravity seems rather impressive.}
as to how fast the charge should delocalize as we adjust the parameter 
$\Delta$.  Thus, it
is worth thinking for a moment about how to build a quantitative
measure of the delocalization in a black hole solution.   A natural
choice is to decompose the electric field using spherical harmonics, to
measure the dipole, quadrapole, and higher moments of the electric
field and to declare that the charge has delocalized on an angular
scale $\theta \sim 1/l$ when the charge is close enough to the black hole
that the spherical harmonics of order $l$ have become small.
However, to do so, one must introduce a foliation of spacetime
by spheres and an action of the rotation group on those
spheres.  Since, at finite $\Delta$, the spacetime is not
spherically symmetric, there is some ambiguity here.  A
convenient choice is to use the spheres and SO(3) action 
defined by isotropic coordinates centered on the black hole.
A calculation of the delocalization rate is then
straightforward, as our spacetimes are described by the
Majumdar-Papapetrou solutions \cite{MP}, for which the 
fields take a simple form in isotropic coordinates.
The detailed results are not important here, though they will be
for the 10-d supergravity analogues discussed below.

\section{The supergravity version}

The main focus of this exposition is a version of charge delocalization
in 9+1 dimensional supergravity theories.  
Recall that, in addition to black holes, higher dimensional
supergravity has what are known as black {\it brane} solutions.  A black
brane is like a black hole except that its horizon does not have
the topology of a sphere, and is often not compact.  For example, in an
$n$ dimensional spacetime, an event horizon with topology $S^{n-2} \times
{\bf R}$, where the ${\bf R}$ factor
is associated with a null direction, would be
referred to as a black hole, while an event horizon with topology
$S^{n-3} \times {\bf R} \times {\bf R}$ is a black string, and 
$S^{n-2-p} \times {\bf R}^p \times {\bf R}$ is a black $p$-brane.

We will be particularly interested in the class of branes known
as ``D-branes'' in the context of 9+1 supergravity theories.
Here, the label D stands for `Dirichlet' and D in no way denotes
the dimension of the brane.  Thus, we will often refer to
D$p$-branes; 
e.g., a D1-brane is a string of the ``Dirichlet'' type.
For a discussion of what the term ``Dirichlet'' means in this 
context, see \cite{JP}.  Below, we focus on
the extremal limits of these solutions\footnote{Unfortunately, 
the extremal limits.  Instead, they are
singular static solutions, with the norm of the Killing field
vanishing at the singularity. The singularity is null for $p \le 5$ 
but timelike for $p \ge 6$.  }.  
The associated supergravity
solutions containing only a one brane have
a single length scale, $r_p \propto Q_p^{1/(7-p)}$, where $Q_p$ is
the charge of the brane. 

A relevant fact about D$p$-branes is that branes of different dimensions
(different values of $p$) carry different kinds of charge.  Thus, 
even in the extremal limit, one cannot in general construct static
solutions containing D$p$-branes with different values of $p$, as
the gravitational attraction is not balanced by the electrostatic
repulsion.  However, there is also a dilaton field in the supergravity
theories of interest, and this produces a repulsive force
between two branes even when their dimensions differ.  For the right
combinations of extremal branes, it can be arranged for these dilatonic
forces to hold the branes apart, so that exactly static solutions
can in fact be constructed.  A class of examples on which we
will focus here consists of static spacetimes containing D$p$-branes
and D$(p-4)$-branes.  
Roughly speaking, static solutions
exist whenever the D$(p-4)$-branes are `parallel' to the D$p$-branes.
That is, the final solutions have a $(p-4)$ dimensional set of
commuting spacelike Killing vector fields\footnote{In fact, the symmetry
group contains the Poincare' group of $(p-4)+1$ dimensional
Minkowski space.}.  
We will focus on such systems from now on and, as
a result, we can refer to the D$p$-brane simply as `the big brane'
and the D$(p-4)$-brane simply as `the little brane' or `the smaller
brane.'  
As before, we may consider a family of such
solutions labeled by a parameter $\Delta$ which describes the separation
of the two branes.

Let us now think of the larger ($p$-dimensional) D-brane as the analogue
of the black hole in section II, and let us think of the smaller
($(p-4)$-dimensional) brane as the analogue of the charge.  We may 
again ask if the charge of the smaller brane `spreads out over the 
larger brane' as the separation $\Delta$ goes to zero.  If it does not, 
then by taking the limit $\Delta \rightarrow 0$, one could form
`hairy' brane solutions which could not be completely characterized
by the amount of big- and little-brane charge they carried.
Instead, they would also require the specification of the {\it distribution}
of the little-brane charge over the big brane.

Now, it is not a priori clear what our 3+1 Einstein-Maxwell
intuition should
tell us about the delocalization of little-brane charge in the current
context.  We now consider to a different gravitating theory and,
while the objects
being described are similar to black holes, they are singular.
It turns out that it is again possible to study
delocalization by direct calculation
as the analogues of the Majumdar-Papapetrou metrics can be found
for an arbitrary separation $\Delta$
between the branes.  Such solutions were studied in
\cite{SM} for the case of $p=5$, and
more generally in \cite{MaPe}.  The final result is that 
the little-brane delocalizes for the cases $p=4$ and $p=5$,
but not for $p=6$.   That is, 
for $p=4,5$ the limiting soliton as $\Delta \rightarrow 0$ in fact has 
a $p$-dimensional group of translation symmetries in addition to the rotational
symmetries that one would expect of a $p$-brane. However, 
for $p=6$, the only new symmetries of the limiting solution are
the rotational ones.  Corresponding localized solutions were in
fact constructed explicitly in \cite{ITY}
in what is known as the ``near-core limit.''
One might say that, for $p=6$, the $2$-brane charge spreads out in the angular
directions around the $6$-brane, but not in the translational directions along
the $6$-brane.  It is probably not a coincidence that D$4$-branes and
D$5$-branes
have (naked) null singularities, while the D$6$-brane in fact has
a naked timelike
singularity.    Following a common practice, we will not 
discuss cases with $p \ge 7$.
The reason for this is that such solutions have a more complicated
structure at infinity.  Since our branes live in a (9+1)-dimensional spacetime,
$p=7$ branes behave
like point particles in 2+1 dimensions, producing conical deficit angles
at infinity.  The $p=8$ branes produce fields which do not fall off at infinity, and
the $p=9$ branes extend to infinity in all directions.

One can measure the rate at which charge delocalizes in a manner
similar to that discussed in section II.  Here, we
are most interested in the delocalization in the directions along
the brane.  Thus, 
we can simplify our lives by considering not just a single (fully localized) 
little brane, but a shell of such branes placed in a spherically symmetric
manner about the bigger-brane.  Thus, even at finite $\Delta$, we may consider
solutions that have the symmetries of $S^{8-p}$ in addition to a 
$(p-4)$-dimensional
translational symmetry.  

These spacetimes may be 
equipped with a nice set of coordinates \cite{MaPe} analogous to the
isotropic coordinates of the Majumdar-Papapetrou solutions.  This introduces
a radial function $r$ and an action of the four-dimensional
translation group that moves the little brane 
along the big-brane.  To measure the extent to which little-brane 
charge is delocalized, consider some surface $r=r_0$ outside the
shell of little branes and Fourier transform the 
fields on this
surface with respect to the four-dimensional translation group.
For a shell at $r=\Delta < r_0$, we may say
that the solution has delocalized on a length scale $\lambda$
when the corresponding Fourier component falls below, say, 
$e^{-10}$ times the value it has when the shell is at $r=r_0$.  In fact,
we may take a limit $r_0 \rightarrow \infty$ to make this
definition independent of $r_0$.   
We display the results here so that the reader may properly
appreciate the
comparison with the field theoretic calculations in section IV.
In the limit where the separation $\Delta$ between the
branes is small, the distance scale over which the
solution is
delocalized behaves like $\lambda \sim r_4^{3/2}/\Delta^{1/2}$ for
the case $p=4$, and 
it behaves like $\lambda \sim r_5\sqrt{\ln(r_5/\Delta)}$ for $p=5$, while
the $p=6$ case does not delocalize.

After reading this section,  the traditional relativist may feel
caught up in a swirl of D's, $p$'s, $Q$'s and various other letters of the
alphabet.  Such a reader should take a moment to collect their thoughts, 
as this final section will not be more familiar.  In regard to many
statements below, the traditional relativist may feel unqualified to
just what is `reasonable' or `natural.'  This is, of course, to be
expected with a new subject, and I urge such a reader not to worry
overly much.  The main goal should be to take away a broad
overview of the argument and some appreciation for the results.


\section{Field Theory Duals}


Now, describing charge delocalization in supergravity is all
well and good, but the current excitement in string theory concerns the
so-called Maldacena conjecture.  This conjecture states that
aspects of supergravity
are described by certain quantum field theories, even though those
theories do not include gravity when viewed in the usual way.
For details, see the original papers \cite{Juan,IMSY} or the
recent review \cite{rev}.  Here we content ourselves
with an extremely rough statement of the conjecture, which is that 
supergravity physics near the `horizon' (the locus where the norm of
the static Killing field vanishes) of a D$p$-brane is in fact completely
described by a (non-gravitating) quantum field theory.   The idea is that
simple superravity quantities supergravity may in
principle be quite complicated when written in terms of the gauge theory, but that
nevertheless such a `dictionary' that tells us how to translate supergravity
physics into field theory physics can in fact be constructed.  Furthermore, this
dictionary has the property that classical gravitational effects on the 
supergravity side of this correspondence are mapped to strongly quantum mechanical
effects on the field theory side.  For a discussion of how our 
current context of multiple separated branes fits into the 
Maldacena conjecture, see \cite{MaPe}.

For now, we simply quote a few results that will paint
a backdrop for the connections we wish to make between supergravity and field
theory.  The fact that we are interested in the near-horizon physics of the
D$p$-brane means that we will be considering $SU(N)$ super Yang-Mills theory in
$p+1$ dimensions.  Here, the value of $N$ is related to the charge of the
branes, the string length $l_s$, and the string coupling $g_s$ through 
$r_p = l_s (g_s N)^{1/(7-p)}$.  Now, it is well known that Yang-Mills theory in 4 Euclidean
dimensions contains instantons.  As a result, Yang-Mills theory in $p+1$
dimensions (for $p \ge 4$) 
contains solitons which are just the lift of the instanton solutions to
$p+1$ dimensions.  The result is a $(p-4)$-brane shaped soliton, and such 
solitons
in the Yang-Mills theory are to be associated in the Maldacena conjecture with the
D$(p-4)$-branes of the supergravity theory.

The next part of our story will be to guess what aspect of the Yang-Mills theory
should correspond to the spreading of the charge in the supergravity solution.
This is certainly a guess, as it is not something which can be derived from the
known aspects of the correspondence.  However, we will see that there is an 
extremely natural candidate.
Now, a soliton can be viewed as a coherent `lump' of
classical field that holds itself together through the non-linear dynamics of the field.
It turns out that, in super Yang-Mills theory, the size $\rho$ of this lump has
no preferred value.  Taking any instanton solution to the equations of motion and
scaling it by a constant factor again gives a static
solution. 
Thus, there are static soliton solutions in our $p+1$ Yang-Mills theory with
any value of the scale size $\rho$.  

Furthermore, one may allow $\rho$ to vary 
over the $(p-4)+1$ dimensional world-volume of the soliton.
Such solutions are not static, but 
if the distortions are small and of long wavelength (much larger
than the string length) it turns out that $\rho(x)$ behaves just like a 
massless $(p-4)+1$ dimensional field.  
To be a bit more precise, because 
the soliton can point in roughly $N$ directions in gauge space, $\rho$
can be thought of as a roughly $N$-dimensional vector of massless scalars.
Thus,
$\rho^2$ acts like a sum $\sum_{i=1}^N \phi_i^2$ over massless scalar fields.  
Normalizing the scalar fields $\phi_i$ canonically, one finds 
$\rho^2 = l_s^{(p-3)}g_s \sum_{i=1}^N \phi_i^2$.

The soliton in the Yang-Mills theory gives a nice picture of
a D$(p-4)$-brane sitting inside a D$p$-brane, but one might ask
how to encode a separation of the
two branes into the Yang-Mills theory.  This is done
through another set of fields and, for our purposes, the important property
is that, when these fields take some nonzero value $\Delta$, a mass
scale $m(\Delta)$ is generated which interacts with the 
field $\rho$.  If the D$(p-4)$-brane shell is located at
$r=\Delta$, then this
mass is known to be $m= l_s^{-2} \Delta$, in units where 
$\hbar =1$.  This means that our description of $\rho$ as a free field
is really only valid for wavelengths shorter than some infra-red cutoff 
$\Lambda_{IR} (\Delta) = l_s^2/\Delta$ and for wavelengths longer than
the string scale due to the short-distance cutoff described above.  

So, since $\rho$ tells us how spread out the instanton is in the appropriate directions, 
it is natural to expect that it corresponds, via the
Maldacena conjecture, to the spread of the D$(p-4)$-brane
charge in supergravity, at least when the D$(p-4)$ brane is close to the horizon
of the D$p$-brane.  Now, in order to match the energy
of the extremal supergravity solution, the energy in the
field theory must be that of a ground state.
Let us therefore ask how large $\rho$ should
be in a typical low-energy
quantum state of the gauge theory.  Classically, $\rho$ can be
set to any size we desire without changing the energy.    
Quantum mechanically, an attempt to confine $\rho$ to a small range
of values, minimizing the uncertainty in $\rho$,
requires a large momentum conjugate to $\rho$ and thus a large energy.
Thus, quantum fluctuations will effectively cause the value of $\rho$ to be non-zero
in the ground state.
We can estimate
a scale for the effective size of $\rho$ by computing, for example, 
$\langle \rho^2 \rangle$ for the various cases.  To do so, we use the fact that, between
the appropriate infra-red and ultra-violet cutoff scales, $\rho$ acts much like a free
field.  Outside this range of wavelengths, it is reasonable to assume that 
the fluctuations of $\rho$ are small.  Thus, we can estimate $\langle \rho^2 \rangle$
using the properties of massless scalar fields in various dimensions.  The result is
\begin{eqnarray}
&{\rm For \  p=4},& \ \ \langle \rho^2 \rangle = (l_s g_s) N (\Lambda_{IR}-\Lambda_{UV}), \cr
&{\rm For \  p=5},& \ \ \langle \rho^2 \rangle = (l_s^2 g_s) N \ln(\Lambda_{IR}/\Lambda_{UV}), \cr
&{\rm For \  p=6}, & \ \ \langle \rho^2 \rangle = (l_s^3 g_s) N (\Lambda_{UV}^{-1}-
\Lambda_{IR}^{-1}).
\end{eqnarray}
Converting to the supergravity parameters and
using the near-horizon
limit, these results become

\begin{eqnarray}
\label{YM4}
&{\rm For \  p=4},& \ \ \sqrt{\langle \rho^2 \rangle} \sim  r_4^{3/2} \Delta^{-1/2}, \cr
\label{YM5}
&{\rm For \  p=5},& \ \ \sqrt{\langle \rho^2 \rangle} \sim r_5 \sqrt{\ln (\Delta/r_5)}, \cr
&{\rm For \  p=6},& \ \ \sqrt{\langle \rho^2 \rangle} \sim  r_6^{1/2} l_s^{1/2}.
\end{eqnarray}

Note that for $p=4$ and $p=5$ these agree precisely with the delocalization
rates of the supergravity solutions, while for $p=6$ the field
theory predicts that the delocalization is bounded as $\Delta \rightarrow 0$
by a value proportional to a positive
power of the string scale; i.e., that it is
small compared to the scales of classical supergravity.
As explained in \cite{MaPe},
similar arguments in the field theory tell us that certain other
kinds of charge delocalization should also occur
in supergravity.  Due to their more complicated nature, these
predictions are harder to check directly.  Nevertheless, some
preliminary investigations \cite{GKTM} seem to support these
predictions.
Thus, the phenomenon of charge delocalization provides a new
kind of evidence in support of the Maldacena conjecture.  We hope
that it may also provide new insight into the nature of 
the supergravity/field theory correspondence.

\smallskip

{\centerline {\bf Acknowledgements}}

\smallskip

The author would like to thank Andr\'es Gomberoff, David 
Kastor, Jennie Traschen, Sumati Surya, and especially
Amanda Peet for the pleasure of collaborating with them
on the work which let to this commentary.  Thanks also to Jorma
Louko and Amanda Peet for comments on an earlier draft.  This work was
supported in part by NSF grants PHY94-07194 and  PHY97-22362 and by funds
provided by Syracuse University.


\end{document}